\documentclass[epj]{svjour}
\usepackage{cite}
\usepackage{epsfig}
\usepackage{amssymb}
\usepackage{braket}
\usepackage{here}
\usepackage{color}
\usepackage{graphicx}
\usepackage{epstopdf}
\usepackage{float}
\usepackage{color}
\usepackage{siunitx}
\usepackage{soul,color}
\soulregister\cite7
\soulregister\ref7
\usepackage{ulem}
\usepackage{caption}
\usepackage{hyperref}
\usepackage{array,multirow,makecell} 
\usepackage{caption,tabularx,booktabs}
\usepackage{keyval}
\usepackage{graphicx}
\usepackage{comment}

\begin{document}

\title{Giant dipole resonance  in Sm isotopes within TDHF method
}

\author{A. Ait Ben Mennana\inst{1} \thanks{\emph{e-mail:} azdine.benmenana@gmail.com} \and M. Oulne\inst{1} \thanks{\emph{e-mail:} oulne@uca.ma}}

\institute{\inst{1} High Energy Physics and Astrophysics Laboratory, Department of Physics,
	Faculty of Sciences SEMLALIA, Cadi Ayyad University,
	P.O.B. 2390, Marrakesh, Morocco.}

\date{Received: date / Revised version: date}

\abstract{
In this work, we have studied the isovector giant dipole resonance (IVGDR) in even-even Sm isotopes within  time-dependent Hartree-Fock (TDHF) with four Skyrme forces SLy6, SVbas, SLy5 and UNEDF1. The approach we have followed is somewhat similar to the one we did in our previous work in the region of Neodymium (Nd, Z=60) [\href{https://iopscience.iop.org/article/10.1088/1402-4896/ab73d8}{Physica Scripta (2020)}]. We have calculated the dipole strength of $ ^{128-164}\text{Sm}$,  and compared with the available experimental data.  An overall   agreement  between them is obtained. The dipole strength in neutron-deficient $ ^{128-142}\text{Sm}$ and in neutron-rich $^{156-164}\text{Sm}$ isotopes are predicted. Shape phase transition as well as shape coexistence in Sm isotopes are also investigated in the light of IVGDR. In addition, the correlation between the quadrupole deformation parameter $\beta_{2}$ and the splitting  $\Delta E/ \bar{E}_{m}$ of the giant dipole resonance (GDR) spectra is studied. The results confirm that $\Delta E/ \bar{E}_{m}$ is proportional to quadrupole deformation $\beta_{2}$. 
}
\maketitle
\section{Introduction}
\qquad Giant resonances (GRs) represent an excellent example of collective modes of many,if not all, particles in the nucleus~\cite{harakeh2001}. GRs are of particular importance because they currently provide the most reliable  information about the bulk behavior of the nuclear many-body system. The so-called isovector giant diople resonance (IVGDR) is the oldest and best known of giant resonances. This is due to high selectivity for isovector $ E_{1} $ in photo-absorption experiments. Several attempts of theoretical description of GDR have been made using the liquid drop model. Among them, Goldhaber and Teller (GT) interpreted it as collective vibrations  of the protons moving against the neutrons in the nucleus  with the centroid energy of the form $ E_{c}  \propto  A^{-1/6}$\cite{goldhaber1948}. Somewhat later, Steinwedel and Jensen (SJ)  interpreted it as a vibration of proton fluid against neutron fluid with a fixed surface where the centroid energy has the form 
$ E_{c}  \propto  A^{-1/3}$~\cite{speth1981}. The experimental data are adjusted by a combination of these two \cite{berman1975}: in light nuclei, the data follow the law $ A^{-1/6} $, while the dependence $ A^{-1/3} $ becomes more and more dominant for increasing values of A. Since its first observation \cite{bothe1937}, it has been much studied  both experimentally (see for example Refs.\cite{carlos1971,carlos1974,berman1975,donaldson2018}) and theoretically (see for example Refs.\cite{goeke1982,maruhn2005,reinhard2008,yoshida2011,benmenana2020}).\\

The GDR spectra of  nucleus can predict its shape (spherical, prolate, oblate, triaxial). It has a single peak for heavier spherical nuclei while in light nuclei it is split into several fragments \cite{harakeh2001}. In deformed nuclei, the GDR strength is split in two components corresponding to oscillations of neutrons versus protons along and perpendicular to the symmetry axis \cite{speth1991,harakeh2001}.
Several microscopic approaches have been  employed to study GDRs in  deformed nuclei  such as  Separable Random-Phase-Approximation (SRPA) \cite{reinhard2008,reinhard2007c}, time-dependent Skyrme-Hartree-Fock method ~\cite{maruhn2005,fracasso2012}, Relativistic Quasi-particle Random Phase Approximation (RQRPA) \cite{ring2009} and  Extended Quantum Molecular Dynamics (EQMD) \cite{wang2017}. Experimentally, the GDR  is induced by various ways such as photoabsorption~\cite{carlos1971,carlos1974, Masur2006} inelastic scattering \cite{donaldson2018, ramakrishnan1996},$\gamma$-decay~\cite{gundlach1990}. \\

 The time-dependent Hartree-Fock (TDHF)~\cite{dirac1930}  method has been employed in many works to investigate GRs in nuclei. It  provides a good approximation for GR. Early, TDHF calculations concentrated on giant monopole resonance (GMR)\cite{blocki1979,chomaz1987} because they  require only a spherical one-dimensional code. In the last few years with the increase in computer power, large scale TDHF calculations become possible with no assumptions on the spatial symmetry of the system\cite{maruhn2005,maruhn2006,stevenson2004}. Such calculations are performed by codes using a fully three dimensional (3D) Cartesian grid in coordinate space \cite{sky3d}.\\

In our previous work~\cite{benmenana2020}, TDHF method provided an accurate description of the GDR in $^{124-160}\text{Nd}$ isotopes. Four Skyrme forces were used in this work. We obtained an overall agreement with experiment with slight advantage for SLy6~\cite{CHABANAT1998}. In this paper, we aim to study another  even-even isotopic chain namely $^{128-164}\text{Sm}$ with four Skyrme forces SLy6~\cite{CHABANAT1998}, SLy5~\cite{CHABANAT1998}, SVbas\cite{reinhard2009} and UNEDF1~\cite{kortelainen2012}. The first three forces were used in our previous work~\cite{benmenana2020} and gave  acceptable results for GDR in Nd isotopes. The new Skyrme force UNEDF1 provided also satisfactory results in.
Many previous experimental and theoretical works have studied the isotopic chain of Samarium Sm (Z = 62). From the experimental point of view one can see for example Ref.\cite{ carlos1974}) and from the theoretical one Refs.\cite{ yoshida2011, wang2017}. Besides the study of GDR, many works (Refs.\cite{ yoshida2011, ring2009,tao2013}) studied the so-called pygmy dipole resonance (PDR) which correspond to low-energy E$ _{1} $ strength in nuclei with a pronounced neutron exces. The pygmy mode  is regarded as vibration of the weakly bound neutron skin of the neutron-rich nucleus against the isospin-symmetric core composed of neutrons and protons \cite{paar2007}. In Ref. \cite{yoshida2011}, the authors studied PDR in some spherical nuclei such as  $^{144}\text{Sm}$ and deformed ones such as $^{152-154}\text{Sm}$. For spherical nuclei, they found a concentration of the E$ _{1} $ strength in low-energy between 8 and 10 MeV, whereas for deformed nuclei the dipole strength is fragmented into low-energy states. They also showed that the nuclear deformation increases the low-lying strength E$ _{1} $ at E $<$ 10 MeV. The PDR mode  is out of our current work in which we aim at a description of the GDR which lie  at a high excitation energy range of  $\sim$ 10-20 MeV.\\

In this paper, the TDHF approximation~\cite{negele1982} has been applied to study the GDR and shape evolution in even-even Sm (Z=62) isotopes from mass number A=128 to A=164. This study is done with  SKY3D code~\cite{sky3d} which  uses a fully three dimensional (3D) Cartesian grid in coordinate space  with no spatial symmetry restrictions and includes all time-odd terms. Consequently, it is possible to study both spherical and deformed system within the limitation of mean field theory. Due to the open-shell nature of these nuclei,
pairing and deformation properties must be taken into account in this study.
 Firstly, a static calculation  gives some properties of the ground-state of the nucleus like root mean square (r.m.s), $\beta_{2}$, $\gamma$. In dynamic calculation, the ground-state of the nucleus is boosted by imposing a dipole excitation  to obtain the GDR spectra and some of its properties (resonance energies, width). \\   

The paper is organized as follows: in Sec.\ref{sec2}, we give a brief description of  TDHF method and the GDR in deformed nuclei. In Sec.\ref{sec3}, we present  details of the numerical calculations. Our results and discussion are presented in Sec.\ref{sec4}. Finally, Sec.\ref{sec5} gives the summary.

\section{Time-Dependent Hartree-Fock method (TDHF) to giant resonances}\label{sec2}

\subsection{TDHF method}
\qquad The time-dependent Hartree-Fock (TDHF) approximation has been extensively discussed in several references \cite{engel1975,kerman1976,koonin1977}. A brief
introduction of the TDHF method is presented as follows.\\
\qquad The TDHF is a self-consistent mean field (SCMF) theory which was proposed by Dirac in 1930 \cite{dirac1930}. It generalizes the static hartree-Fock (HF) and has been very successful in describing the dynamic properties of nuclei such as for example, giant resonances  \cite{maruhn2005,stevenson2004,reinhard2007,blocki1979} and Heavy-ion collisions \cite{simenel2018,maruhn2006}.\\
 \qquad The TDHF equations are determined from the variation of Dirac action 
\textit{\begin{equation}{\label{eq1}}
S \equiv S_{t_0,t_1}[\psi] = \int_{t_0}^{t_1} dt  \bra{\psi(t)} \bigg(i\hbar\frac{d }{dt} - \hat{H} \bigg) \ket{\psi(t)}, 
\end{equation}}
where $ \ket{\psi} $ is the Slater determinant, $ t_{0} $ and $ t_{1} $ define the time interval, where the action S is stationary between the fixed endpoints $ t_{0} $ and $ t_{1} $, and $ \hat{H}  $ is the Hamiltonian of the system. The energy of the system is defined as  $ E = \bra{\psi} \hat{H} \ket{\psi} $, and we have 
\textit{\begin{equation}{\label{eq2}}
\bra{\psi}\frac{d }{dt}\ket{\psi} = \sum_{i=1}^{N} \bra{\varphi_{i}}\frac{d }{dt}\ket{\varphi_{i}},
\end{equation}}
where \textit{$ \ket{\varphi_{i}} $} are the occupied single-particle states.
The action S can be expressed as

\begin{eqnarray}{\label{eq3}}
S & = & \int_{t_0}^{t_1} dt \bigg( i\hbar \sum_{i=1}^{N} \bra{\varphi_{i}}\frac{d }{dt}\ket{\varphi_{i}} - E[\varphi_{i}]\bigg) \nonumber \\
& = & \int_{t_0}^{t_1} dt \bigg( i\hbar \sum_{i=1}^{N} \int dx \,\varphi_{i}^{*}(x,t) \frac{d }{dt}\varphi_{i}(x,t) - E[\varphi_{i}] \bigg)
\end{eqnarray}
The variation of the action S
with respect to the wave functions $ \varphi_{i}^* $ reads
\begin{equation}{\label{eq4}}
\frac{\delta S}{\delta \varphi_{i}^*(x,t)} = 0,
\end{equation}
for each $ i = 1....N  $, $ t_{0}\leq t \leq {t_{1}} $ and for all $ x $. More details can be found  for example in Refs.~\cite{kerman1976, simenel2012}. We finally get  the TDHF equation 
\begin{equation}{\label{eq5}}
i\hbar\frac{\partial }{\partial t}\varphi_{i}(t) = \hat{h}[\rho(t)]\varphi_{i}(t) \quad \text{for} \quad 1\leq i \leq \text{N}.
\end{equation}
where $ \hat{h} $ is the single-particle Hartree-Fock Hamiltonian.

The TDHF equations (\ref{eq5}) are solved \textit{iteratively} by a small time step $ \Delta t $  during which we assume that the Hamiltonian remains constant. To conserve the total energy E, it is necessary to apply a symmetric algorithm by time reversal, and therefore to estimate the Hamiltonian at time $ t + \frac{\Delta t}{2} $ to evolve the system between time $ t \;\text{and}\; t+ \Delta t$ \cite{flocard1978,bonche1976}
\begin{equation}{\label{eq6}}
\ket{\varphi(t+\Delta t)} \simeq e^{-i\frac{\Delta t}{\hbar}\hat{h}(t+\frac{\Delta t}{2})}\ket{\varphi(t)}.
\end{equation} 
\subsection{ Giant dipole resonance in deformed nuclei}
\qquad In deformed axially symmetric nuclei, one of the most spectacular properties of the GDR is its splitting into two components associated to vibrations of neutrons against protons along (K=0) and perpendicularly to (K=1) the symmetry axis. Therefore, the GDR strength represents a superposition of two resonances with energies \textit{$E_{i} \sim R_{i}^{-1} \sim A^{-1/3} $} \cite{speth1981} where \textit{R} is the nuclear radius,  and even three resonances in the case of asymmetric nuclei. This splitting has been observed experimentally \cite{carlos1974,berman1975,Masur2006, donaldson2018} and treated theoretically by different models \cite{maruhn2005,reinhard2008,yoshida2011}.  For the axially symmetric prolate nuclei, the GDR spectra present two peaks where the low-energy $E_{z}$ corresponds to the oscillations along the  major axis of symmetry and the high-energy $E_{x}= E_{y}$ corresponds to the oscillations along  transverse minor axes of the nuclear ellipsoid, due to $E \sim R^{-1}$. For an oblate nucleus, it is the opposite situation to the prolate case. For  triaxial nuclei, the oscillations along three axes are different ,\textit{i.e}., $ E_{x}\neq E_{y}\neq E_{z}$. For  spherical nuclei, the vibrations along three axes degenerate and their energies coincide $E_{x} =  E_{y} = E_{z} $.

\section{Details of Calculations}\label{sec3}
\qquad In this work, the  GDR in  even-even  $^{128-164}\text{Sm}$  isotopes has been studied by using the code Sky3D (v1.1)~\cite{sky3d} . This code solves the HF as well as TDHF equations for Skyrme interactions \cite{SKYRME1958}. Calculations  were performed with four Skyrme functional: SLy6 \cite{CHABANAT1998}, SLy5\cite{CHABANAT1998}, SVbas\cite{reinhard2009}, UNEDF1~\cite{kortelainen2012}. These Skyrme forces are widely used for the ground state properties (binding energies, radii...) and dynamics (as giant resonances) of nuclei including deformed ones. In particular they provide a reasonable description of the GDR: SLy6\cite{maruhn2005,reinhard2008}, SVbas\cite{reinhard2009}, SLy5\cite{fracasso2012} and UNEDF1~\cite{kortelainen2012}.
The parameters set of these functionals used in this study is shown in Table~\ref{tab1}.
\begin{table}[!htbp]
	\centering
	\caption{Parameters ($ t,x $) of the  Skyrme forces used in this work. \label{tab1}}
	{\begin{tabular}{@{}ccccc@{}} \hline
			Parameters & UNEDF1 & SVbas &  SLy6 & SLy5 \\
			\hline
			$ t_{0} $ (MeV.fm$^3$) & -2078.328 & -1879.640 &  -2479.500 & -2484.880\\
			$t_{1}$ (MeV.fm$^5$) & 239.401 & 313.749 &  -1762.880 & 483.130\\
			$ t_{2} $ (MeV.fm$^5$)& 1574.243 & 112.676  & -448.610 & -549.400\\
			\quad	$ t_{3} $ (MeV.fm$^{3+3\sigma}$)& 14263.646 & 12527.389 &  13673.000 & 13763.000\\
			$ x_{0} $ & 0.054 & 0.258 &  0.825 & 0.778\\
			$ x_{1} $ & -5.078 & -0.381 &  -0.465 & -0.328\\
			$ x_{2} $ & -1.366 & -2.823 &  -1.000 & -1.000\\
			$ x_{3} $ & -0.161 & 0.123 &  1.355 & 1.267\\
			$ \sigma $ & 0.270 & 0.300 &  0.166 & 0.166\\
			W$ _{0} $ (MeV.fm$^5$)& 76.736 & 124.633 &  122.000 & 126.000\\
			\hline
	\end{tabular}}
\end{table}

A first step of calculation concerns a static calculation which allows to determine the ground state for a given nucleus. This state is obtained by solving the static HF + BCS equations (\ref{eq8})  in a three-dimensional (3D) Cartesian mesh with a damped gradient iteration method on an equidistant grid and without symmetry restrictions~\cite{sky3d}.

	\begin{equation}{\label{eq7}}
	\hat{h}\psi_{i}(x)= \epsilon_{i} \psi_{i}(x) \quad \text{for} \quad i=1,....,A,
	\end{equation}
	where $ \hat{h} $ is the single-particle Hamiltonien, and $ \epsilon_{i} $ is the single-particle energy of the state  $ \psi_{i}(x) $ with $ x=(\vec{r},\sigma,\tau) $.\\
	We used a cubic box with size a = 24 fm and a grid spacing of $\Delta x$ = 1.00 fm in 	each direction. In SKY3D code \cite{sky3d}, the static HF + BCS equations (\ref{eq7}) are solved 	iteratively until a convergence is obtained ,\textit{i.e}., when for example the sum of the single-particle energy fluctuations becomes less than a certain value determined at the beginning of the static calculation. In this study we take as a convergence value 	$ 10^{-5} $ which is sufficient for heavy nuclei (for more details see Ref.~\cite{sky3d}. The pairing is treated in the static calculation, which allows  to calculate the pairing energy  
	\begin{eqnarray}{\label{eq8}}
	E_{pair}  = \frac{1}{4} \sum_{q \in \{p,n\}} V_{pair,q}\int d^{3}r \vert \xi_{q} \vert^{2} F(r)
	\end{eqnarray}
	where the pairing density $ \xi_{q} $ reads~\cite{sky3d}
	\begin{eqnarray}{\label{eq9}}
	\xi(\vec{r})  =  \sum_{\alpha \in \{p,n\}} \sum_{s} u_{\alpha}v_{\alpha} \vert \psi_{\alpha}(\vec{r},s) \vert^{2}
	\end{eqnarray}
	 where $ v_{\alpha} $, $ u_{\alpha} = \sqrt{1-v_{\alpha}^{2}} $  are the occupation and non-occupation amplitude of single-particle state $ \psi_{\alpha} $ , respectively, and the function $ F=1 $ or $ F = 1 - \rho/\rho_{0} $  gives a pure $\delta$-interaction (DI), also called volume pairing (VDI) where $\rho_{0} \rightarrow \infty$ or density dependent $\delta$-interaction (DDDI), respectively, while 	$ \rho_{0}=0.16$ fm$ ^{-3} $  is the saturation density. $ V_{P,N} $ represents the  pairing strength  which is obtained from the force definition in the SKY3D code~\cite{sky3d}.

	\qquad  In dynamic calculations, the ground-state wave function obtained by the static calculations is excited by an instantaneous initial dipole boost operator in order to put the nucleus in the dipole mode \cite{maruhn2005,simenel2009,stevenson2008}. 
	\begin{equation}{\label{eq10}}
	\varphi_{\alpha}^{(g.s)}(r) \longrightarrow \varphi_{\alpha}(r,t=0)=\exp(ib\hat{D})\varphi_{\alpha}^{(g.s)}(r),
	\end{equation}
	where $ \varphi_{\alpha}^{(g.s)}(r) $ represents the ground-state of nucleus before the boost, b is the boost amplitude of the studied mode , and $ \hat{D} $ the associated operator. In our case, $ \hat{D} $ represents the isovector dipole operator  defined as
	\begin{eqnarray}{\label{eq11}}
	\hat{D} & = & \frac{NZ}{A} \bigg( \frac{1}{Z}\sum_{p=1}^{Z}\vec{z}_p - \frac{1}{N}\sum_{n=1}^{N}\vec{z}_n \bigg) \nonumber \\
	& = & \frac{NZ}{A}\bigg( \vec{R}_Z - \vec{R}_N \bigg),
	\end{eqnarray}
	where $ \vec{R}_Z $ (resp. $ \vec{R}_N $ ) measures the proton (resp. neutron)  average position on the z axis.
	
	 The spectral distribution of the isovector dipole strength is obtained by applying a boost (\ref{eq10}) with a small value of the amplitude of the boost b to stay well in the linear regime of the excitation. For a long enough time,  the dipole moment $ \hat{D} = \bra{\psi(t)}\hat{D}\ket{\psi(t)} $ is recorded along the dynamical evolution. Finally, the dipole strength $ S_{D}(\omega) $ can be obtained by performing the Fourier transform $ D(\omega) $ of the signal $ \hat{D}(t) $, defined as \cite{ring1980}
	\begin{eqnarray}{\label{eq12}}
	S_{D}(\omega) & = & \sum_{\nu} \delta(E-E_{\nu}) \big| \bra{\nu}\hat{D}\ket{0}\big|^2. 
	\end{eqnarray}
	
	Some filtering is necessary to avoid artifacts in the spectra obtained by catting the signal at a certain final time, in order to the signal vanishes at the end of the simulation time. In practice we use windowing in the time domain by damping the signal $ D(t) $  at the final time with $ cos \big(\frac{\pi t}{2T_{fin}}\big)^n $ \cite{sky3d}.
	\begin{equation}{\label{eq13}}
	D(t) \longrightarrow D_{fil} = D(t). cos \bigg(\frac{\pi t}{2T_{fin}}\bigg)^n,
	\end{equation}
	where n represents the strength of filtering and $ T_{fin} $ is the final time  of the simulation. More details can be founded in Refs. \cite{sky3d, reinhard2006} .\\
	In this work, all dynamic calculations were performed in a cubic space with 24 x 24 x 24 fm$ ^3 $ according to the three directions (x, y, z) and a grid spacing of  1 fm. We chose nt= 4000 as a number of time steps to be run, and dt = 0.2 fm/c is the time step, so T$ _ {f}$ = 800 fm/c is the final time of simulation. Pairing is frozen in the dynamic calculation ,\textit{i.e}., the BCS occupation numbers are frozen at their initial values during time evolution.
\section{Results and Discussion}\label{sec4}
\qquad In this section we present our numerical results of static calculations concerning some properties of the ground-state, and dynamic calculations concerning some properties of the GDR for $ ^{128-164}\text{Sm}$ nuclei.
\subsection{Ground-state properties}

\begin{figure}[!htbp]
	\begin{center}
		\includegraphics[scale=0.6]{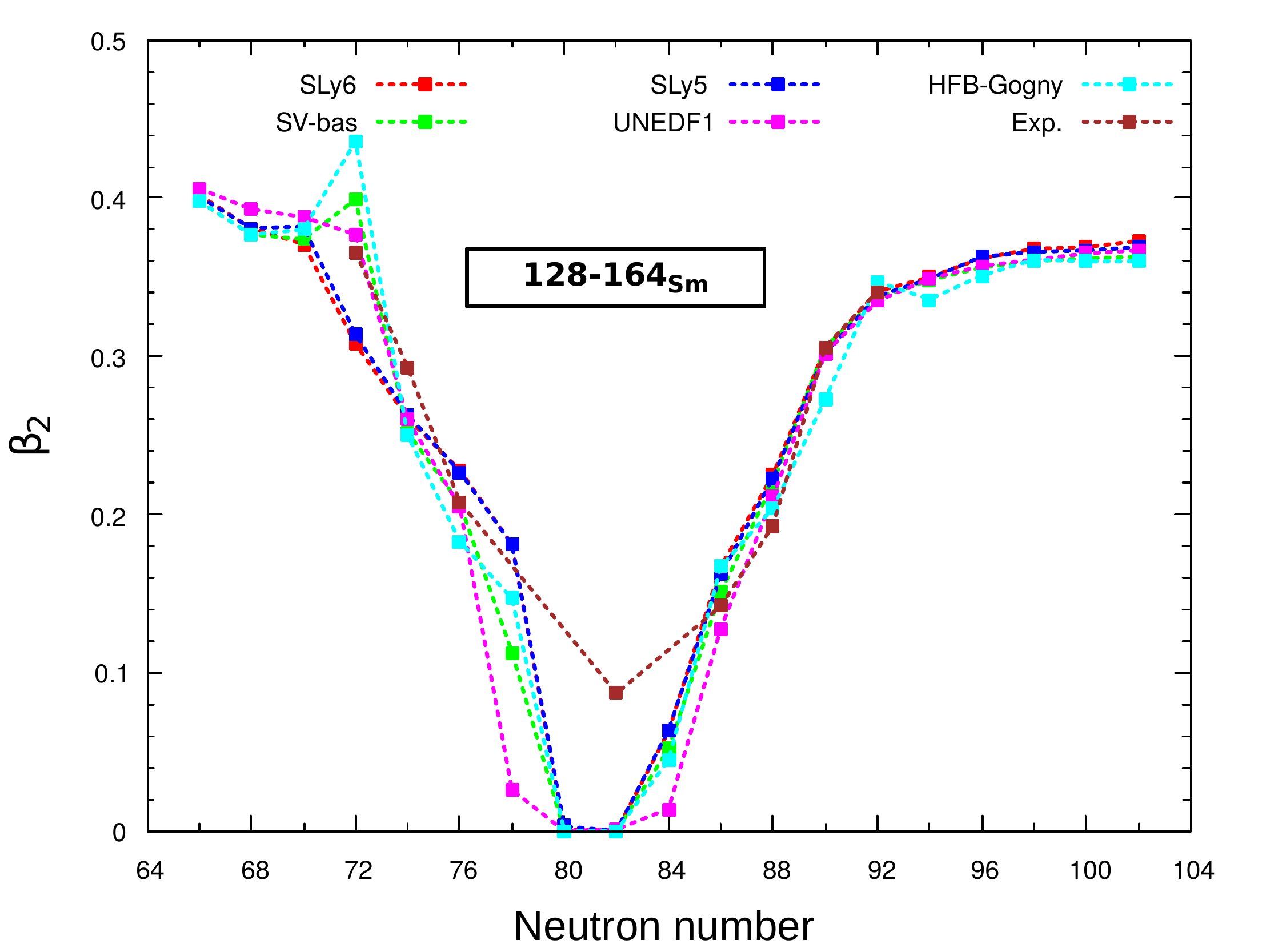}
		\caption{The Quadrupole deformation parameter $ \beta_{2} $ of $ ^{124-160}\text{Nd} $ isotopes as function of their neutron number N. The experimental data are from Ref.  \cite{raman2001}.}
		\label{b2-n}
	\end{center}
\end{figure}

\qquad The isotopic chain of  Sm (Z=62)  studied in this work displays a transition from spherical, when neutron number N is close to magic number N = 82, to the axially deformed shapes  when N increases or decreases \cite{carlos1974,meng2005,wang2017,naz2018}. Among the properties of the ground-state of  nuclei, there are the deformation parameters $\beta_{2}$ and $\gamma$ which give an idea on the shape of the nucleus \cite{ring1980,takigawa2017}. These deformation parameters are defined as follows \cite{sky3d}
\begin{equation}{\label{eq14}}
\beta = \sqrt{a_{0}^2 + 2a_{2}^2} \qquad , \quad \gamma = atan\bigg(\frac{\sqrt{2}a_{2}}{a_{0}}\bigg) 
\end{equation} 

 \begin{equation}{\label{eq15}}
 	a_{m} = \frac{4\pi}{5}\frac{Q_{2m}}{AR^2} \qquad , \quad  R = 1.2  A^{1/3} (fm),
 \end{equation}
 where $ Q_{2m} $ is the quadrupole moment defined as 
 \begin{equation}{\label{eq16}}
 	Q_{2m} = \int \rho(\vec{r}) r^{2} Y_{2m}(\theta,\varphi) d\vec{r} 
 \end{equation}
 The deformation parameters ($\beta$,$\gamma$)  often called Bohr-Mottelson parameters are treated as a probe to select the ground-state of all nuclei in this article. Table \ref{tab2} displays the numerical results obtained for the  deformation parameters ($\beta_{2}$,$\gamma$) based on Eq.~(\ref{eq14}) of $ ^{128-164}\text{Sm} $  isotopes with four Skyrme forces, including the available experimental data from Ref.\cite{raman2001} and the HFB calculations based on the D1S Gogny force~\cite{HFB} for comparison. Fig.~\ref{b2-n} shows the variation of $ \beta_{2} $ as a function of neutrons number N. 

\begin{table}[!htbp]
	\centering
	\caption{The  deformation parameters ($\beta_{2}$,$\gamma$) calculated with UNEDF1, SVbas,  SLy6, and SLy5 are compared with the experimental data are from Ref.\cite{raman2001}, and data from Ref.\cite{HFB}. \label{tab2} }
	{\begin{tabular}{@{}ccccccc@{}} \hline
			Nuclei & UNEDF1 & SVbas &  SLy6 & SLy5 & HFB$\_$Gogny.\cite{HFB}&Exp. \cite{raman2001}\\
			\hline
			$ ^{128}\text{Sm} $ & (0.406; $0.0^\circ$) & (0.398; $4.8^\circ$) &  (0.402; $8.6^\circ$) & (0.401; $7.6^\circ$) & (0.398; $8.0^\circ$)&-----\\
			$ ^{130}\text{Sm} $ &  (0.393; $0.1^\circ$) & (0.377; $0.0^\circ$) &  (0.381; $0.0^\circ$) & (0.381; $0.0^\circ$) & (0.377; $0.0^\circ$)&-----\\
			$ ^{132}\text{Sm} $ & (0.388; $0.0^\circ$) & (0.374; $0.0^\circ$) &  (0.371; $0.0^\circ$) & (0.382; $0.0^\circ$) & (0.380; $0.0^\circ$)&-----\\
			$ ^{134}\text{Sm} $ & (0.377; $0.0^\circ$) & (0.399; $0.0^\circ$) &  (0.308; $14.8^\circ$) & (0.314; $12.2^\circ$) & (0.436; $0.0^\circ$)&0.366\\
			$ ^{136}\text{Sm} $ & (0.260; $21.3^\circ$) & (0.252; $22.5^\circ$) &  (0.261; $22.4^\circ$) & (0.263; $21.9^\circ$) & (0.252; $22.0^\circ$)&0.293\\
			$ ^{138}\text{Sm} $ & (0.205; $27.5^\circ$) & (0.207; $26.1^\circ$) &  (0.228; $25.6^\circ$) & (0.227; $25.2^\circ$) & (0.183; $25.0^\circ$)&0.208\\
			$ ^{140}\text{Sm} $ & (0.026; $14.8^\circ$) & (0.113; $0.0^\circ$) &  (0.181; $27.7^\circ$) & (0.181; $27.6^\circ$) & (0.147; $35.0^\circ$)&-----\\
			$ ^{142}\text{Sm} $ & (0.000; $20.6^\circ$) & (0.000; $14.4^\circ$) &  (0.001; $0.0^\circ$) & (0.003; $17.0^\circ$) & (0.000; $0.0^\circ$)&-----\\
			$ ^{144}\text{Sm} $ & (0.001; $4.0^\circ$) & (0.000; $8.5^\circ$) &  (0.000; $12.7^\circ$) & (0.000; $1.5^\circ$) & (0.000; $0.0^\circ$)&0.087\\
			$ ^{146}\text{Sm} $ & (0.014; $58.0^\circ$) & (0.052; $1.2^\circ$) &  (0.063; $0.0^\circ$) & (0.064; $0.7^\circ$) & (0.045; $2.0^\circ$)&-----\\
			$ ^{148}\text{Sm} $ & (0.128; $0.0^\circ$) & (0.151; $0.2^\circ$) &  (0.167; $3.6^\circ$) & (0.162; $0.0^\circ$) & (0.167; $0.0^\circ$)&0.142\\
			$ ^{150}\text{Sm} $ & (0.211; $0.0^\circ$) & (0.220; $0.0^\circ$) &  (0.225; $0.0^\circ$) & (0.223; $0.0^\circ$) & (0.204; $0.0^\circ$)&0.193\\
			$ ^{152}\text{Sm} $ & (0.302; $0.0^\circ$) & (0.306; $0.0^\circ$) &  (0.305; $0.0^\circ$) & (0.302; $0.0^\circ$) & (0.273; $0.0^\circ$)&0.306\\
			$ ^{154}\text{Sm} $ & (0.335; $0.0^\circ$) & (0.337; $0.0^\circ$) &  (0.341; $0.0^\circ$) & (0.338; $0.0^\circ$) & ((0.347; $0.0^\circ$)&0.341\\
			$ ^{156}\text{Sm} $ & (0.349; $0.0^\circ$) &(0.348; $0.0^\circ$) &  (0.350; $0.0^\circ$) & (0.349; $0.0^\circ$) & (0.336; $0.0^\circ$)&-----\\
			$ ^{158}\text{Sm} $ & (0.357; $0.0^\circ$) & (0.356; $0.0^\circ$) &  (0.362; $0.0^\circ$) & (0.363; $0.0^\circ$) & (0.351; $0.0^\circ$)&-----\\
			$ ^{160}\text{Sm} $ & (0.361; $0.0^\circ$) & (0.360; $0.0^\circ$) &  (0.368; $0.0^\circ$) & (0.366; $0.0^\circ$) & (0.361; $0.0^\circ$)&-----\\
			$ ^{162}\text{Sm} $ & (0.365; $0.0^\circ$) & (0.362; $0.0^\circ$) &  (0.369; $0.0^\circ$) & (0.367; $0.0^\circ$) & (0.360; $0.0^\circ$)&-----\\
			$ ^{164}\text{Sm} $ & (0.367; $0.0^\circ$) & (0.363; $0.0^\circ$) &  (0.373; $0.0^\circ$)& (0.369; $0.0^\circ$) & (0.360; $0.0^\circ$)&-----\\
			\hline
	\end{tabular}}
\end{table}
 From Fig.\ref{b2-n}, we can see the $ \beta_{2} $ values of our calculations are generally close to experimental ones~\cite{raman2001}. On the other hand, there is an agreement between our calculations and HFB theory based on the D1S Gogny force\cite{HFB}. In the vicinity of the region where N = 82, the $ \beta_{2} $ values show minima ($ \beta_{2} \simeq 0 $) as expected because all nuclei with the magic number N=82 are spherical. For the  $ ^{140}\text{Sm}$ nucleus, we find different results between the four Skyrme forces in this study. For the Skyrme forces SLy6 and SLy5, $ ^{140}\text{Sm}$ has a triaxial shape ($\gamma \simeq28.0^\circ$). It has a prolate shape for SV-bas ($\gamma=0.0^\circ$), and has an approximate spherical form for UNEDF1 force ($\beta_{2} \simeq0.026$). For comparison, Calculations by M\"{o}ller et al.\cite{moller2008}, based on the finite-range droplet model, predicted the ground state of $ ^{140}\text{Sm}$  nucleus to be triaxial ($\gamma=30.0^\circ$). In table \ref{tab2}, the ($\beta_{2}$,$\gamma$) values obtained in this work as well as those of HFB theory based on the D1S Gogny force\cite{HFB} and avialable experimental data \cite{raman2001} show a shape transition from spherical $ ^{144}\text{Sm}$ (N=82) to deformed shape below and above the magic neutron number N=82. For $ ^{128-144}\text{Sm}$ isotopes below N = 82, the  isotopic
 chains exhibit a transition from prolate ($\gamma=0.0^\circ$) to spherical shape ($\beta_{2} \simeq0.000$) passing through triaxial form ($ 22.0^\circ\leq \gamma \leq 28.0^\circ $) for $ ^{136-140}\text{Sm}$ isotopes, and for neutron number higher than N = 82, both the experimental and theoretical results  show that the prolate deformation increases gradually and then saturates at a value which closes to $\beta_{2}\simeq$ 0.368.

\subsection{Giant dipole resonance in $^{128-164}\text{Sm}$ nuclei}
\qquad Based on the TDHF ground states  for $ ^{128-164}\text{Sm}$  isotopes  obtained in static calculations, we perform dynamic calculation such as GDR in this work to obtain some of its properties as we will see later.
  \subsubsection{The time evolution of the dipole moment $ D_{m}(t) $}
 
\begin{figure}[!htbp]
  	\centering
  	\includegraphics[width=1.0\textwidth]{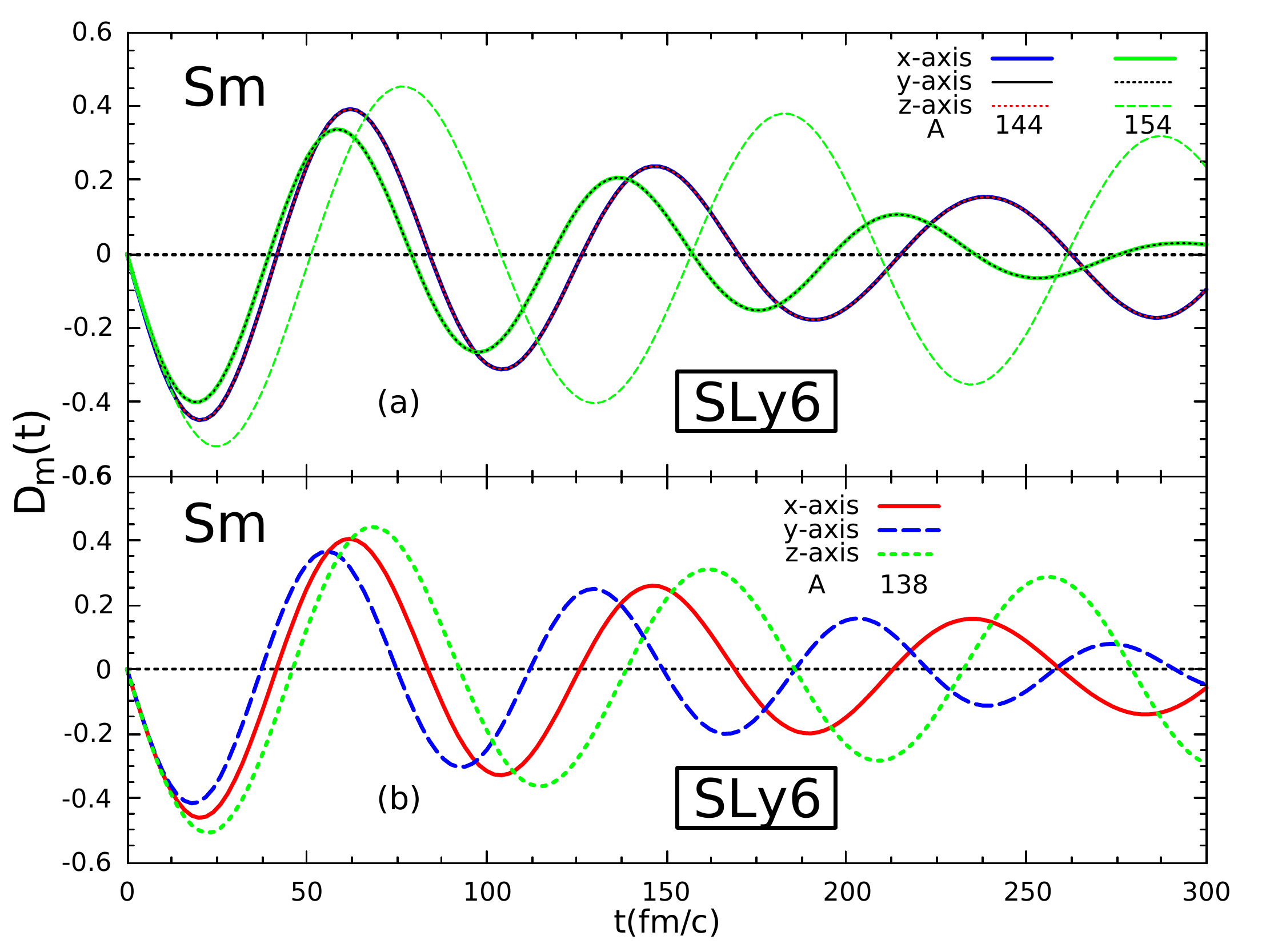} 
  	\caption{The dipole moment $ D_{m}(t) $ as function of the simulation time  t(fm/c)  calculated with the Skyrme force SLy6  for $ ^{138}\text{Sm}$, $ ^{144}\text{Sm}$ and $ ^{154}\text{Sm}$.}
  	\label{fig2}
  \end{figure}

  The dipole moment $ D_{m}(t) $ defined by Eq.~(\ref{eq11}) allows to predict the collective motions of nucleons along the three directions x, y and z.
  The time evolution of $ D^{i}_{m}(t)$ where i denotes x, y and z  of $ ^{138}\text{Sm}$, $ ^{144}\text{Sm}$ and $ ^{154}\text{Sm}$ is plotted in Fig.~\ref{fig2}. We note that the collective motion of nucleons in GDR is  done generally along two axes. The oscillation frequency $\omega_{i}$ is related to the nuclear radius $ R_{i} $ by $\omega_{i} \propto R_{i}^{-1}$ where i$\in$\{x,y,z\}. Fig.~\ref{fig2}(a) shows the time evolution of dipole moment for $ ^{144}\text{Sm}$ and $ ^{154}\text{Sm}$.
  For the $ ^{144}\text{Sm}$ nucleus, the three components $ D^{x}_{m}(t)$, $ D^{y}_{m}(t)$ and $ D^{z}_{m}(t)$ are identical ,\textit{i.e}.,  the oscillation frequencies along the three axes are equal ( $\omega_{x}=\omega_{y}=\omega_{z}$) which confirms that this nucleus has a spherical shape as we predicted in static calculations ($\beta_{2}\simeq0.000$). For the $ ^{154}\text{Sm}$ nucleus, the $ D^{x}_{m}(t)$ and $ D^{y}_{m}(t)$ values are identical and differ from  the values of $ D^{z}_{m}(t)$ ,\textit{i.e}., the oscillation frequencies  along the symmetry z-axis $\omega_{z}$ are lower than that along the two other axes x and y which they are equal $\omega_{x}=\omega_{y}$. This  confirms that $^{154}\text{Sm}$ has a prolate shape because $\omega_{z}< \omega_{x}=\omega_{y}$\cite{Masur2006} which is consistent with our static calculations ($\gamma= 0.0^\circ $). We point out that we found almost the same situation for the prolate nuclei namely $^{130-134}\text{Sm}$ and $^{148-164}\text{Sm}$.
  In Fig.~\ref{fig2}(b), the values of the three components $ D^{i}_{m}(t)$ are different from each other in the case of the $^{138}\text{Sm}$ nucleus. We notice that the oscillation frequencies $\omega_{i}$ along the three axes  are different from each other $\omega_{x}\neq \omega_{y}\neq \omega_{z}$ which confirms that this nucleus has a triaxial shape as we predicted in static calculations ($\gamma \simeq 25.0^\circ $). The same situation occurs for $^{136}\text{Sm}$. We note also that the time evolution of dipole moment $ D_{m}(t) $ is almost the same for the others  Skyrme forces (SLy5, UNEDF1, SVbas) with an exception for some nuclei as $^{140}\text{Sm}$. The periodicity of the three components $ D^{i}_{m}(t)$ allows the  excitation energies $ E_{i} $ to be estimated for the oscillations along each of the three axes. For $^{144}\text{Sm}$, we obtain, for $ D^{x}_{m}(t)$, $ D^{y}_{m}(t)$ and $ D^{z}_{m}(t)$, the same period T $\simeq$ 84.3 fm/c giving an excitation energy $ E_{x}=E_{y}=E_{z}\simeq$ 14.70 MeV. This value is slightly lower than the experimental one $ E_{GDR}^{exp.}$=15.3$\pm$ 0.1~\cite{carlos1974}. The table \ref{tab3} shows the excitation energies for $^{138}\text{Sm}$ and $^{154}\text{Sm}$ nuclei with Skyrme force SLy6. \\
 \begin{table}[!htbp]
 	\centering
 	\caption {The excitation energies along the three axes for $^{138}\text{Sm}$ and $^{154}\text{Sm}$ with Sly6, obtained from the time evolution of $ D^{i}_{m}(t)$.\label{tab3}} 
 	{\begin{tabular}{@{}ccccccccc@{}} \hline
 			Nuclei && $E_{x}$(MeV) &&& $E_{y}$(MeV)&&& $E_{z}$(MeV)  \\
 			\hline
 			$^{138}\text{Sm}$  && 14.75&&&  16.52&&& 13.40\\
 			$^{154}\text{Sm}$   && 15.62&&&  15.62&&&11.90  \\
 			
 			\hline
 	\end{tabular}}
 \end{table}
  

\subsubsection{GDR Spectrum}

The  calculation of  the Fourier transform of the isovector signal D(t) allows to obtain the GDR energy spectrum. The spectral strength  S(E)  (\ref{eq12}) is  simply the imaginary part of the Fourier transform of D(t).

\begin{figure}[!htbp]
 	\begin{center}
 		\begin{minipage}[t]{0.47\textwidth}
 			\includegraphics[scale=0.55]{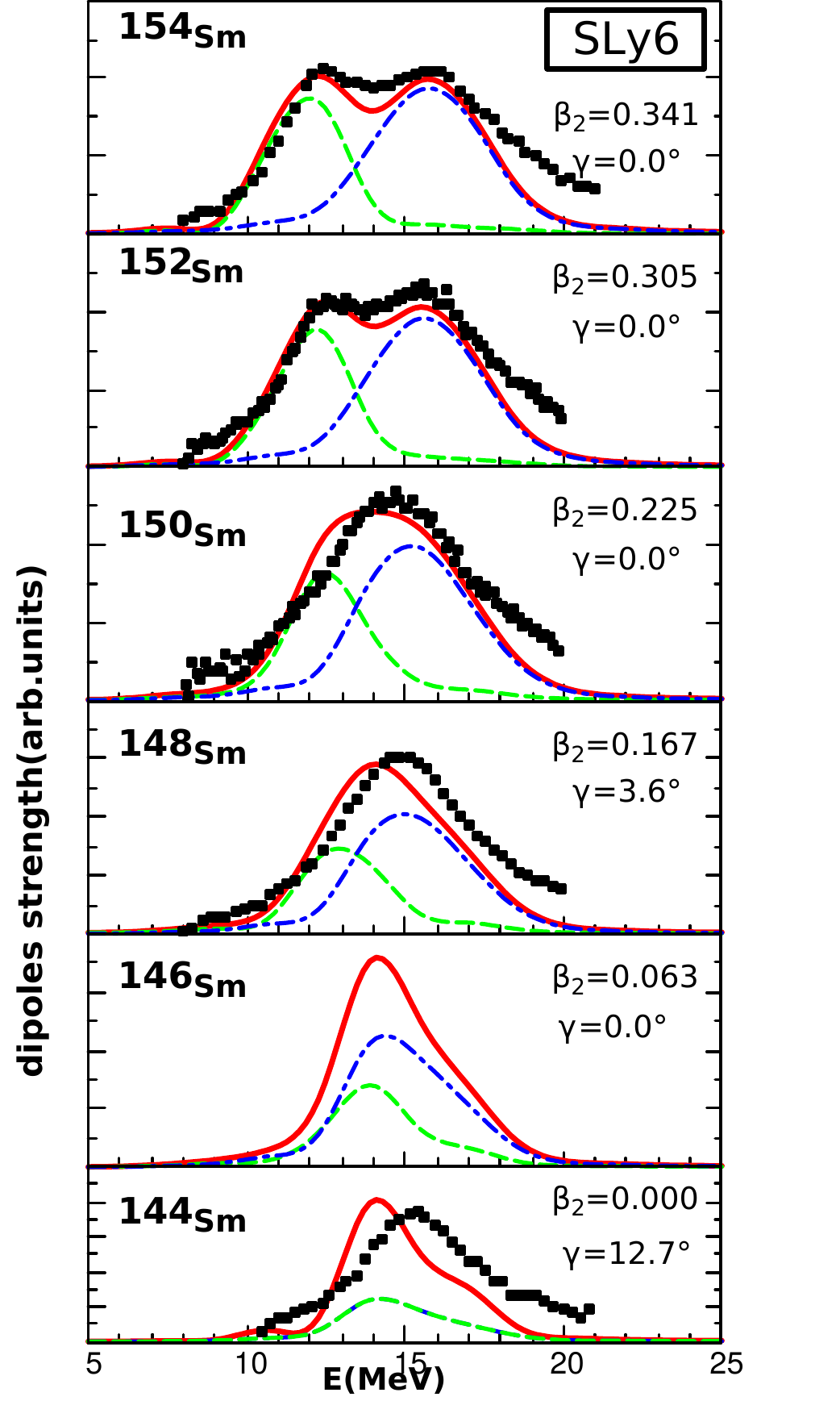}
 			
 		\end{minipage}
 		\vspace{0.5cm}
 		\hspace{0.5cm}
 		\begin{minipage}[t]{0.47\textwidth}
 			\includegraphics[scale=0.55]{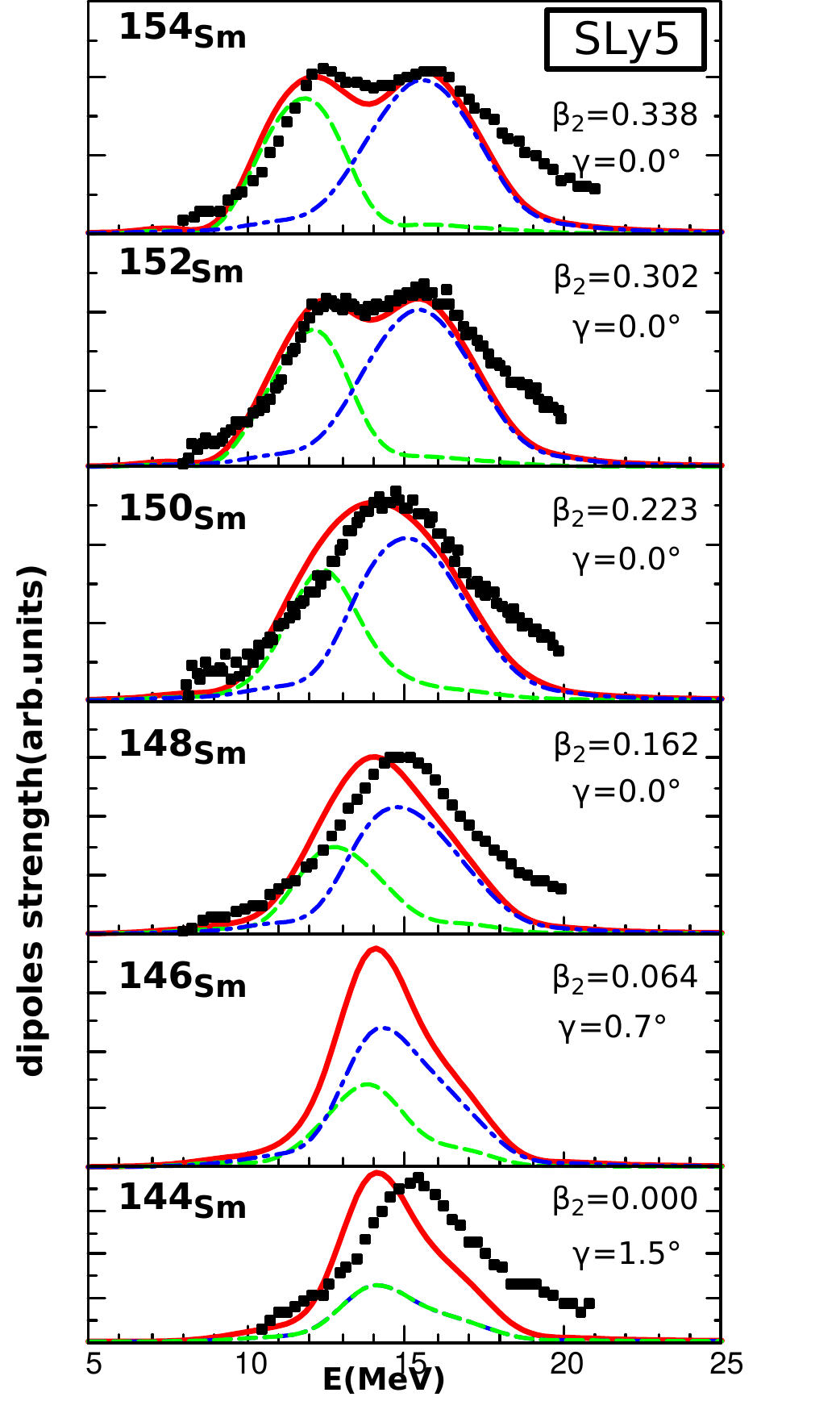}
 		\end{minipage}
 		\begin{minipage}[t]{0.47\textwidth}
 			\includegraphics[scale=0.55]{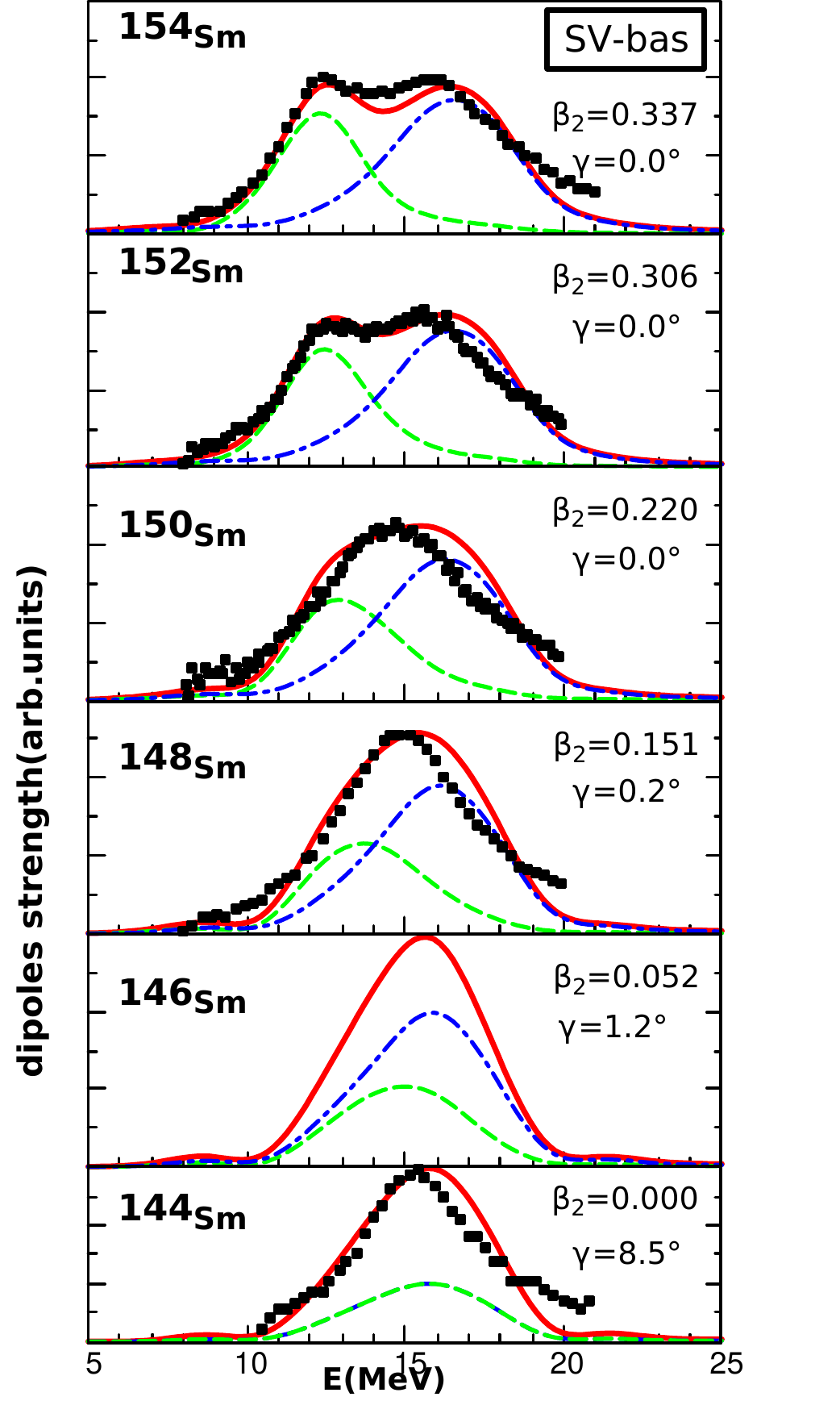}
 		\end{minipage}
 		\hspace{0.5cm}
 		\begin{minipage}[t]{0.47\textwidth}
 			\includegraphics[scale=0.55]{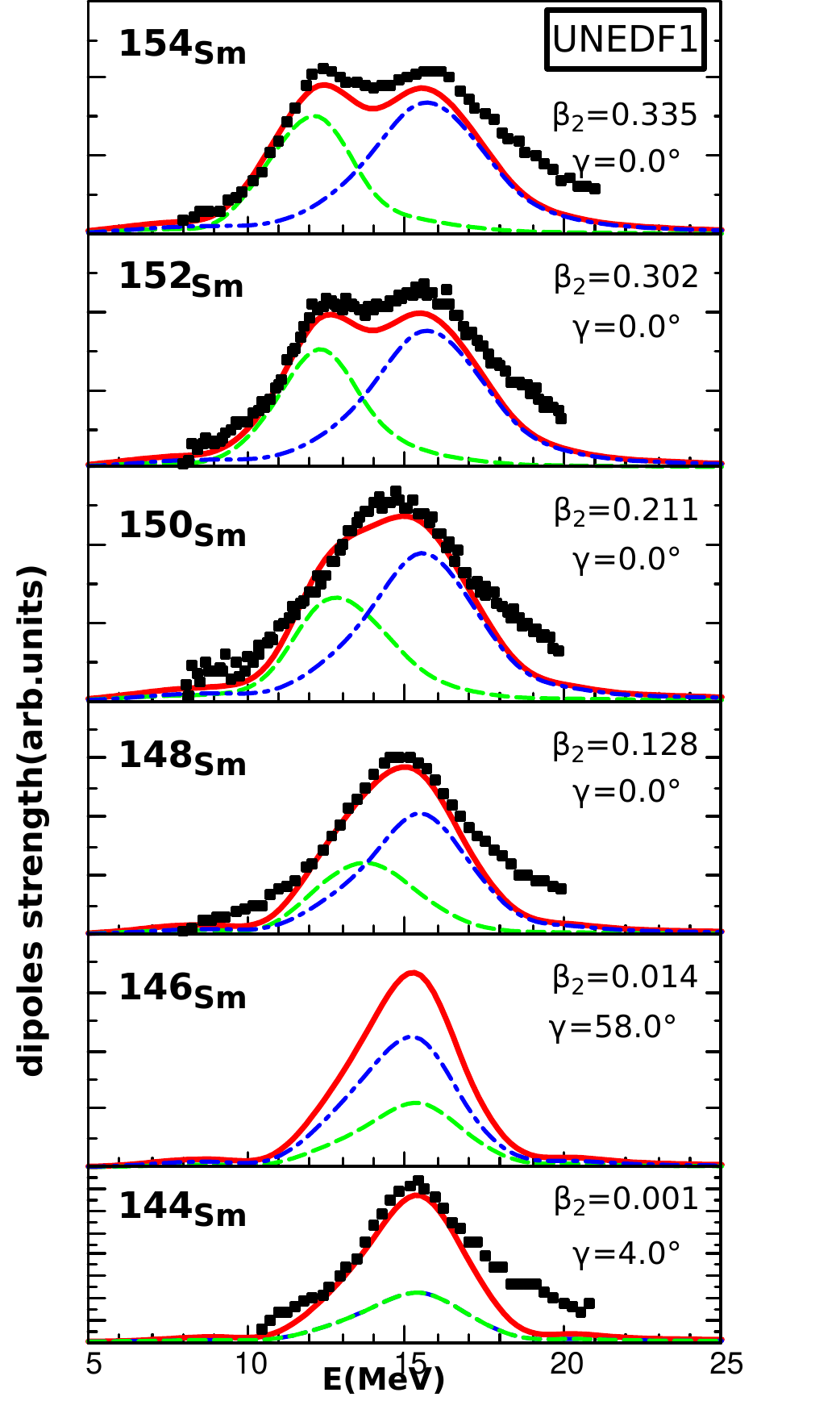}
 		\end{minipage}
 		\caption{(Color online) GDR spectra in the chain of $ ^{144-154}\text{Sm} $ calculated with SLy6, SLy5, SVbas and UNEDF1. The solid(red), dashed(green) and dotted-dashed(blue) lines denote the dipole strengths: total, along the long axis and the short axis (multiplied by 2) respectively. The calculated strength total is compared with the experimental data \cite{carlos1974} depicted by black solid squares.} 
 		\label{fig3}
 	\end{center}	
 \end{figure}

\begin{figure}[!htbp]
 	\centering
 	\includegraphics[scale=0.7]{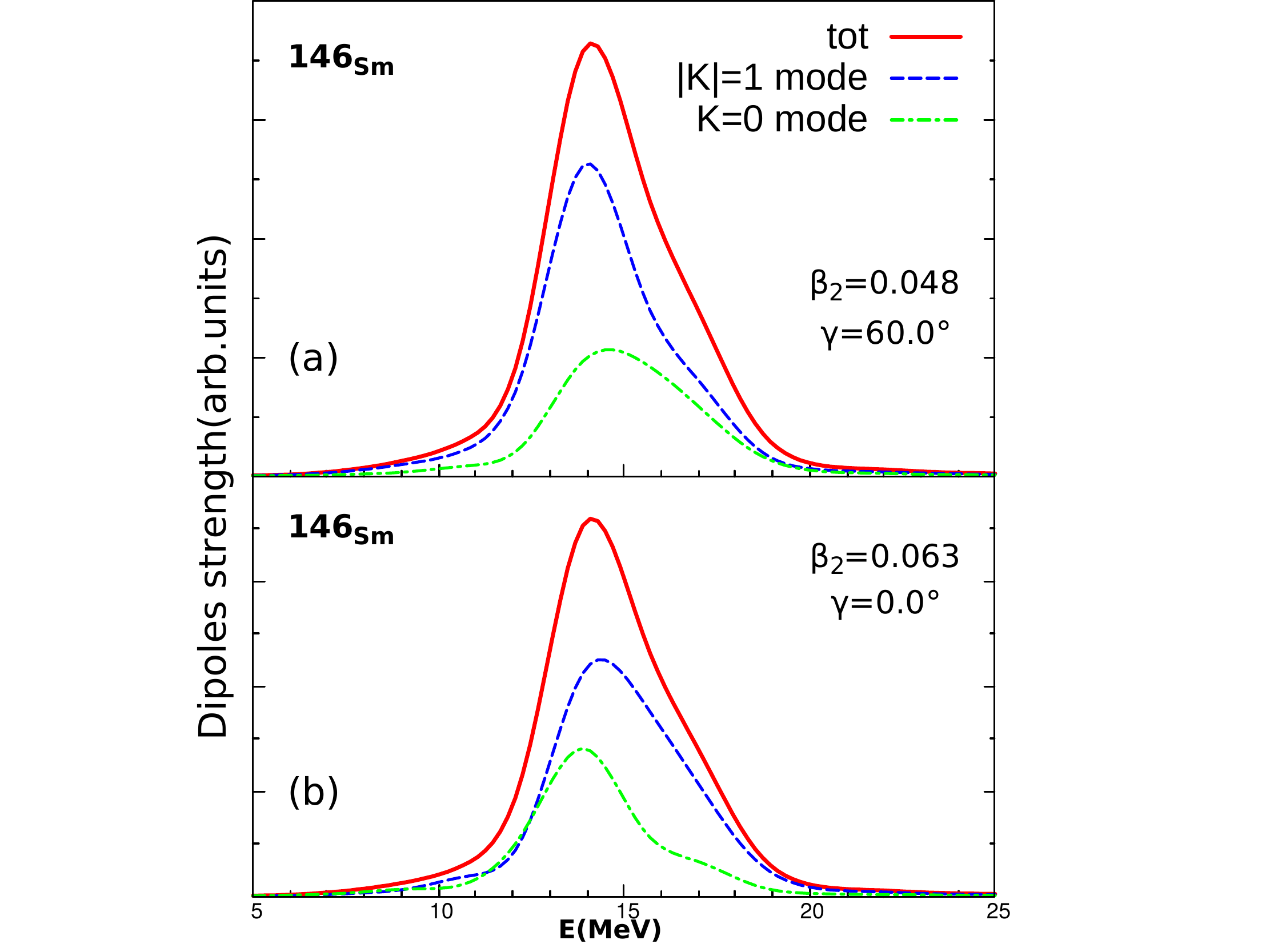}
 	\caption{(Color online) The calculated GDR spectra for $^{146}\text{Sm} $  with the Skyrme force SLy6.} 
 	\label{fig4},
 	
 \end{figure}

\begin{figure}[!htbp]
	\begin{center}
		\begin{minipage}[t]{0.47\textwidth}
			\includegraphics[scale=0.55]{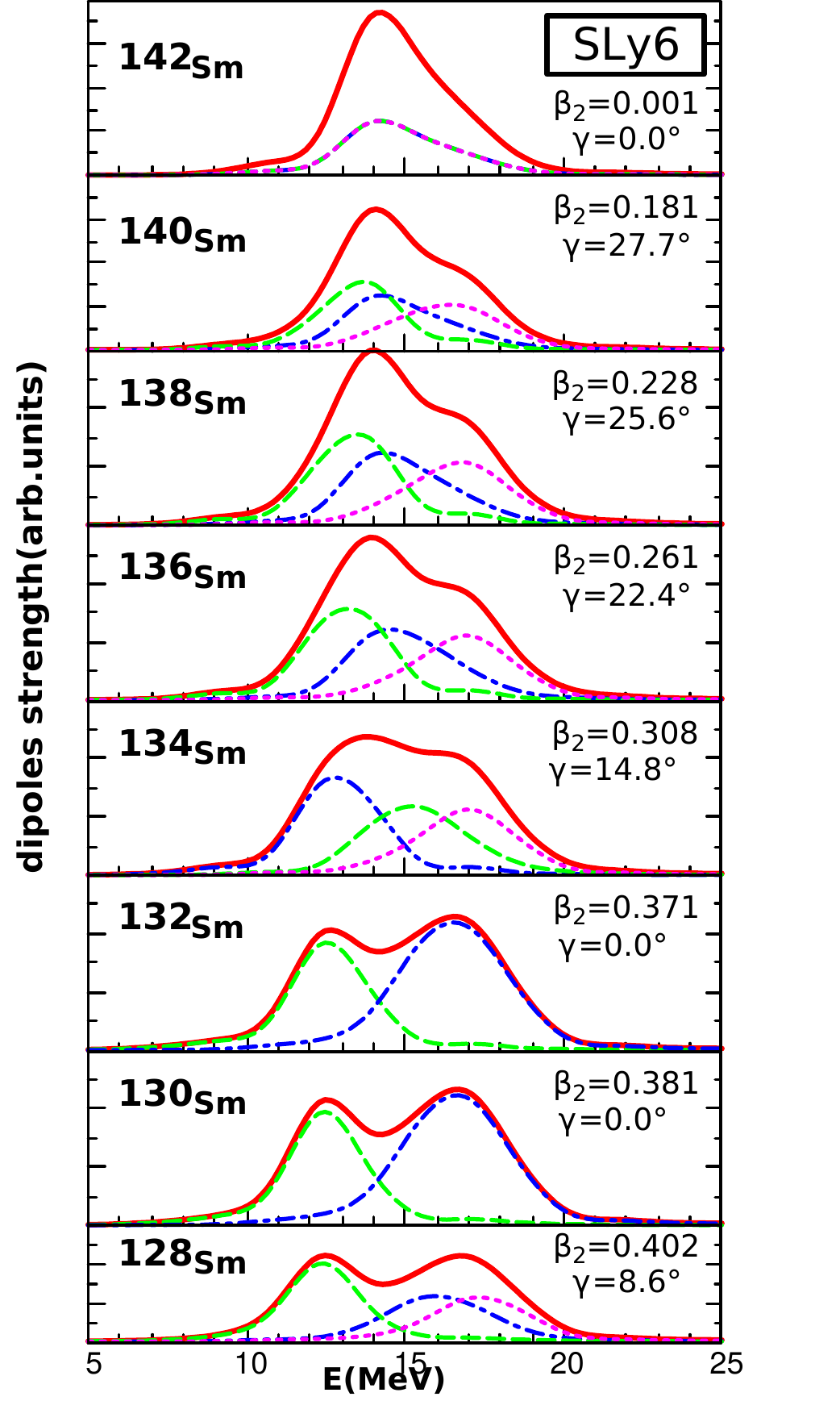}
		\end{minipage}
		\vspace{0.5cm}
		\hspace{0.5cm}
		\begin{minipage}[t]{0.47\textwidth}
			\includegraphics[scale=0.55]{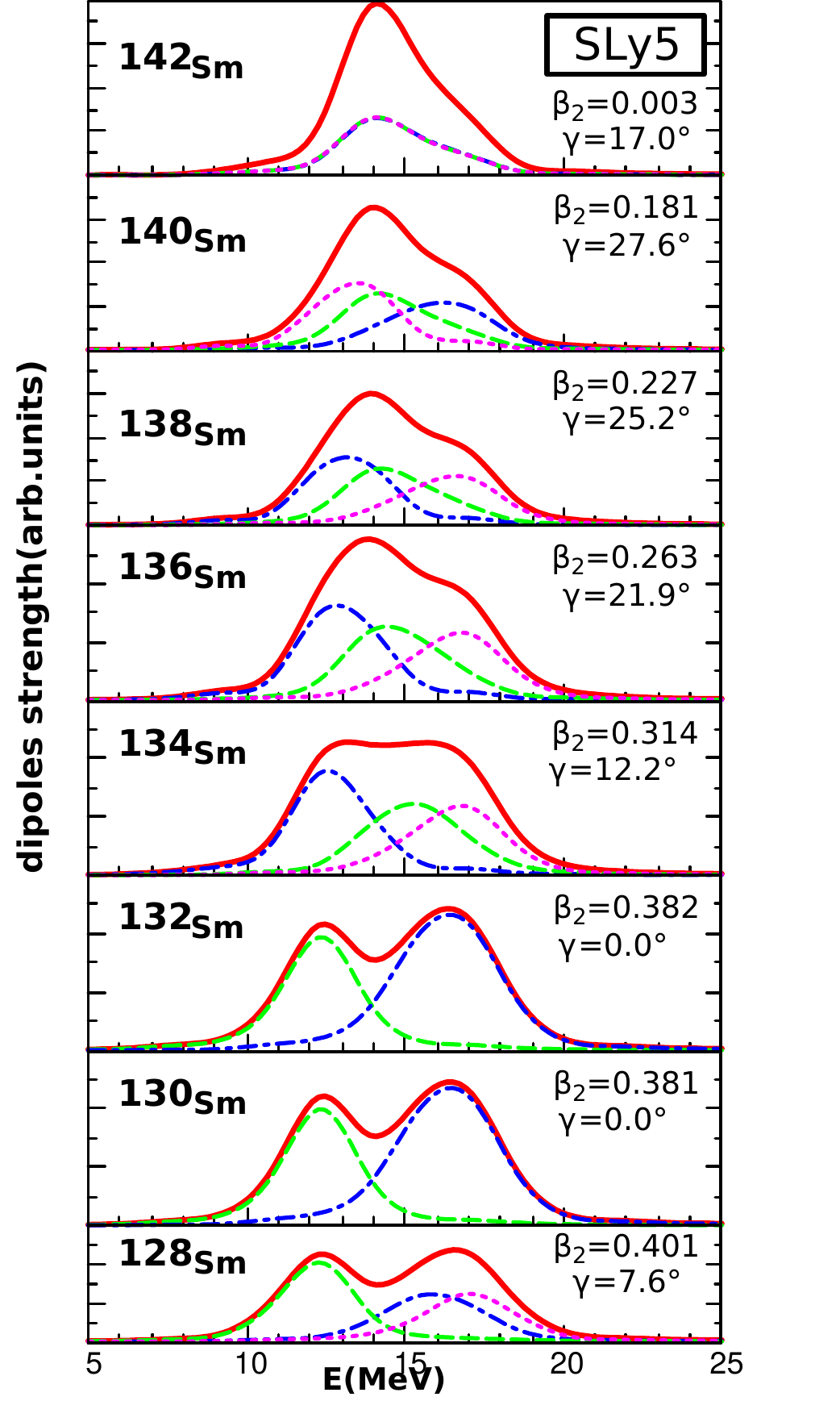}
		\end{minipage}
		\begin{minipage}[t]{0.47\textwidth}
			\includegraphics[scale=0.55]{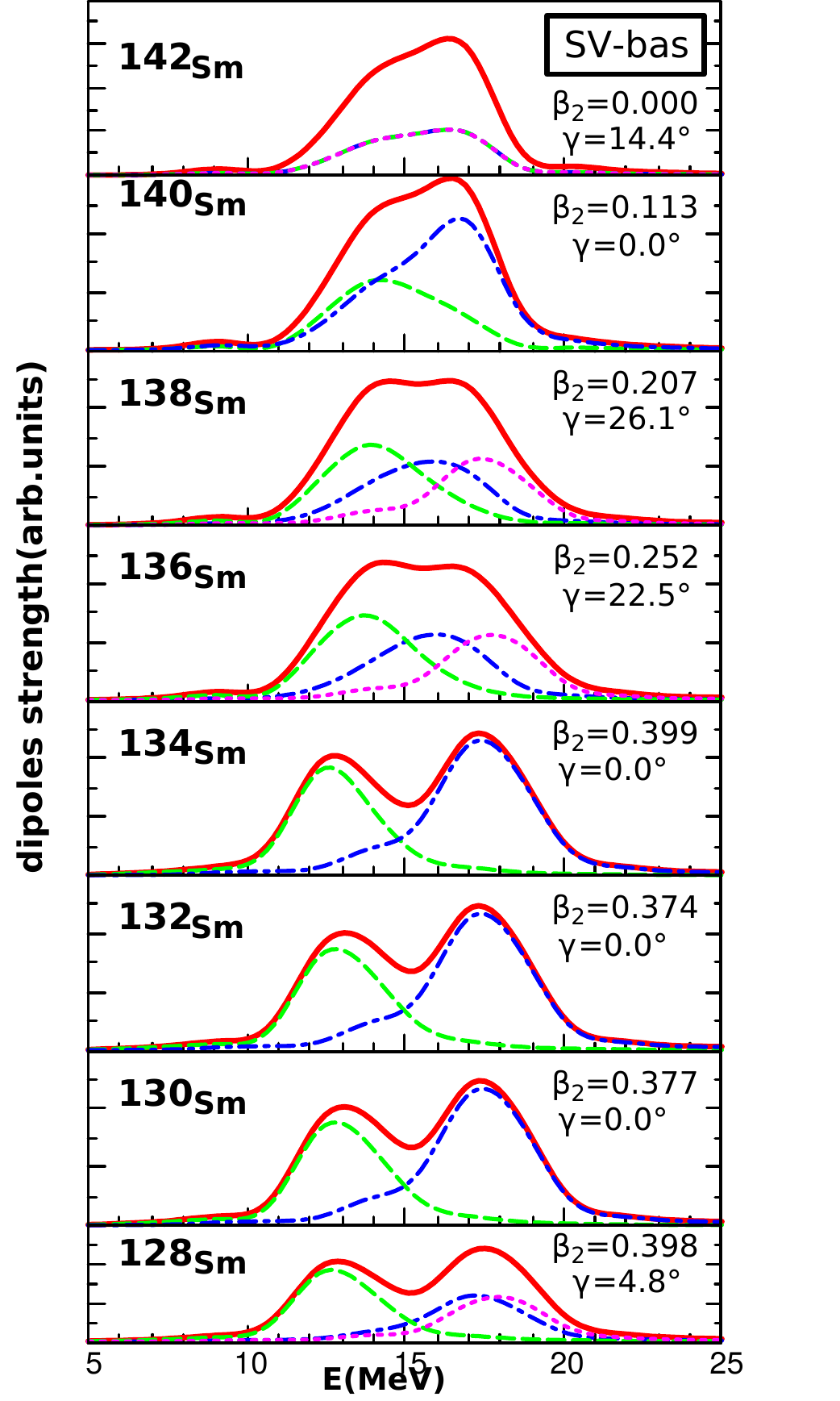}
		\end{minipage}
		\hspace{0.5cm}
		\begin{minipage}[t]{0.47\textwidth}
			\includegraphics[scale=0.55]{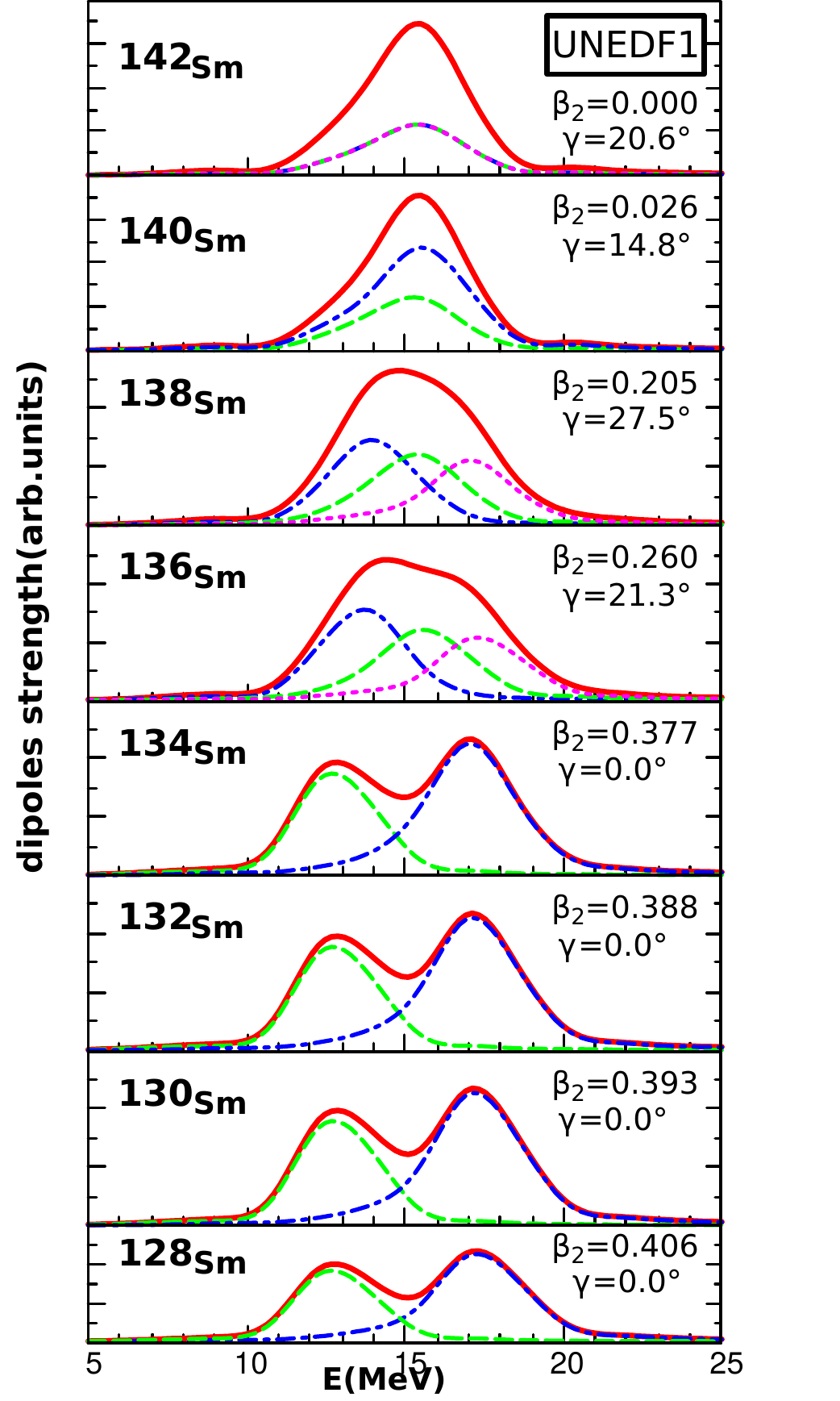}
		\end{minipage}
		\caption{(Color online) GDR spectra in the isotopic chain  $ ^{128-142}\text{Sm} $ calculated with  SLy6, SLy5, SVbas and UNEDF1. The solid(red), dashed(green) and dotted-dashed(blue) lines denote the dipole strengths: total, along the long axis and the short axis(multiplied by 2 except $ ^{136-140}\text{Sm} $) respectively. The dotted (magenta) line denotes the strength along the third middle axis in the case of the triaxial nuclei $ ^{136-140}\text{Sm}$.} 
		\label{fig5}
	\end{center}	
\end{figure}

\begin{figure}[!htbp]
	\begin{center}
		\begin{minipage}[t]{0.47\textwidth}
			\includegraphics[scale=0.55]{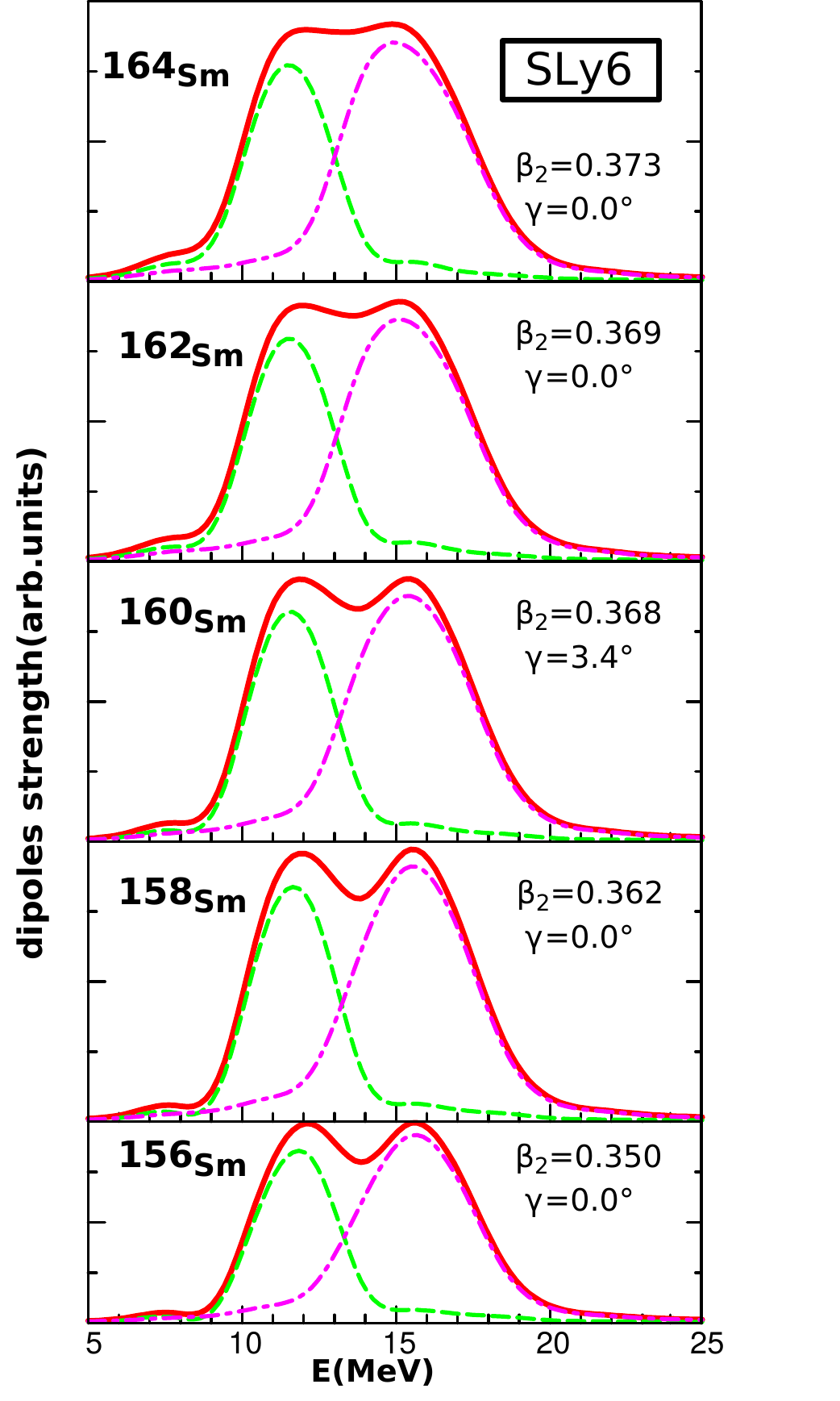}
		\end{minipage}
		\vspace{0.5cm}
		\hspace{0.5cm}
		\begin{minipage}[t]{0.47\textwidth}
			\includegraphics[scale=0.55]{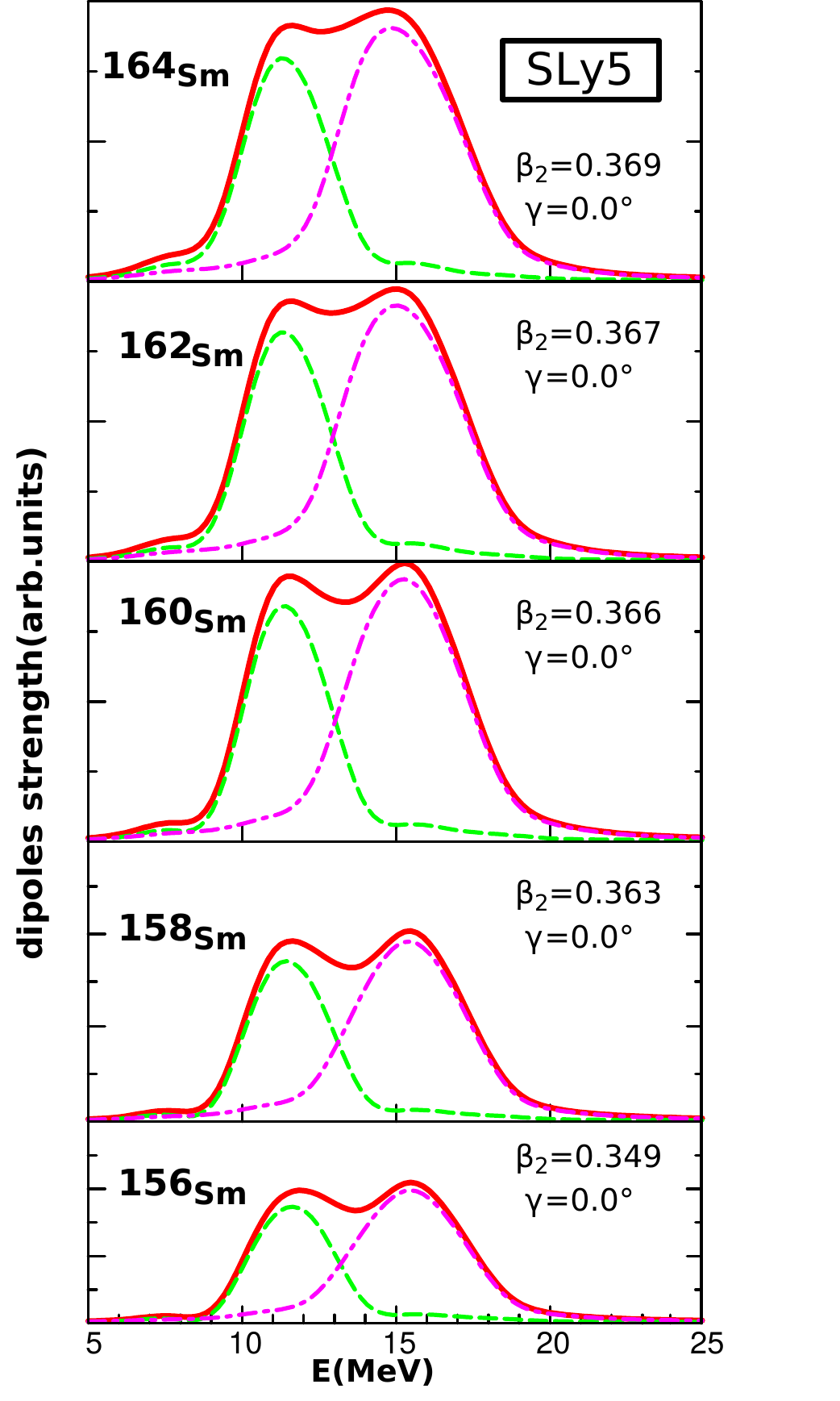}
		\end{minipage}
		\begin{minipage}[t]{0.47\textwidth}
			\includegraphics[scale=0.55]{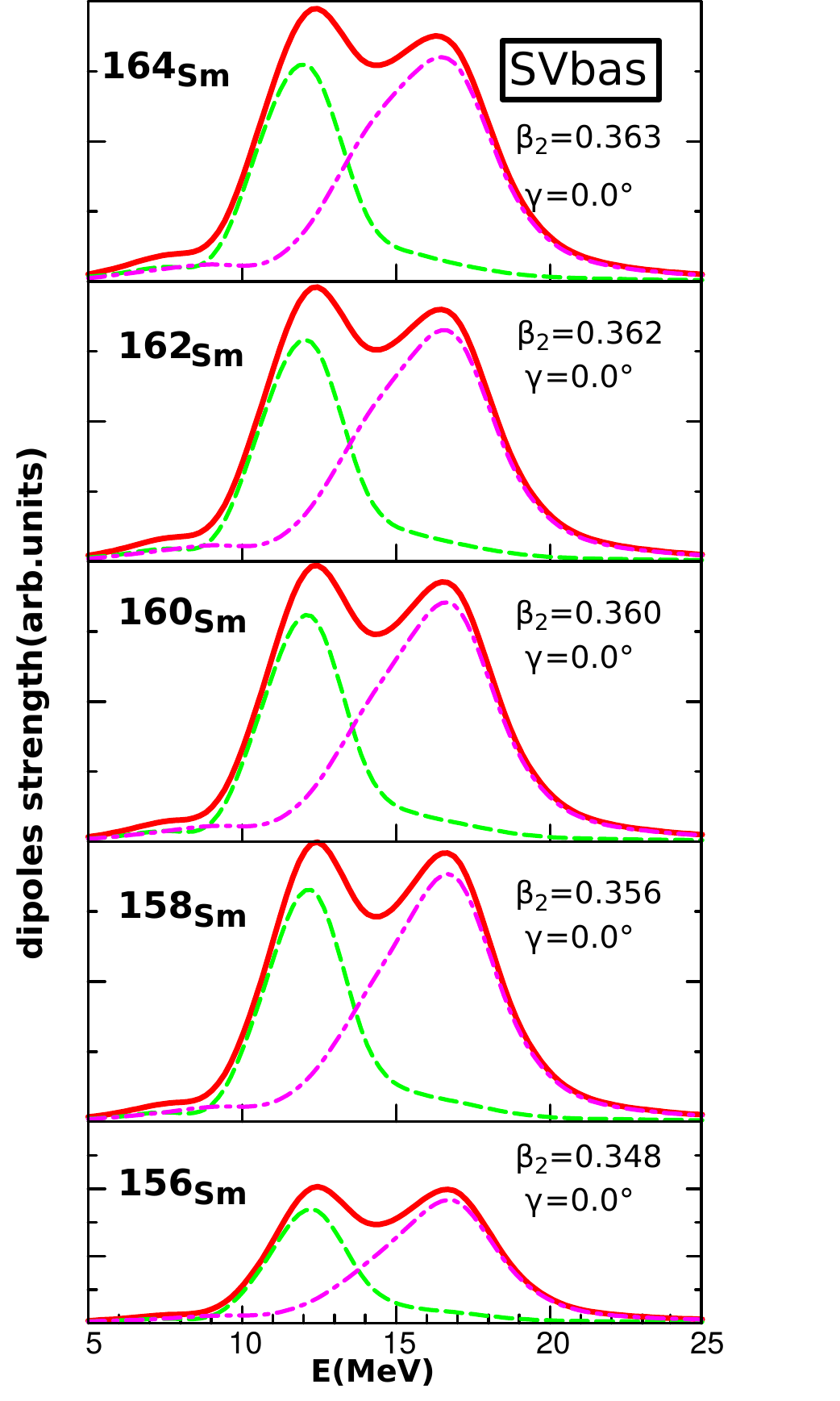}
		\end{minipage}
		\hspace{0.5cm}
		\begin{minipage}[t]{0.47\textwidth}
			\includegraphics[scale=0.55]{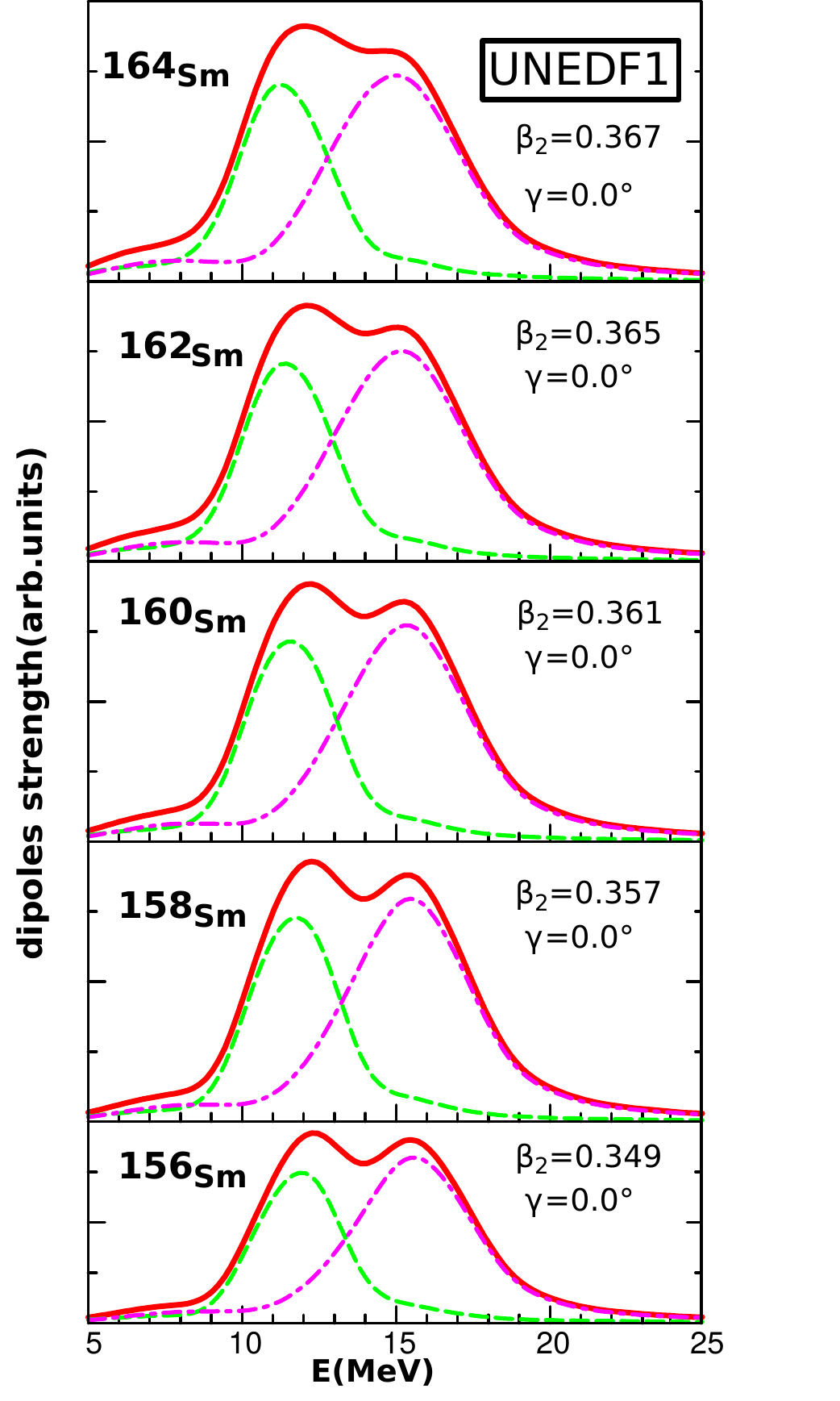}
		\end{minipage}
		\caption{ (color online) The GDR spectra in the isotopic chain  $ ^{156-164}\text{Sm} $ calculated with  SLy6, SLy5, SVbas and UNEDF1. The solid(red), dashed(green) and dotted-dashed(magenta) lines denote the dipole strengths: total, along the long axis and the short axis(multiplied by 2) respectively.} 
		\label{fig6}
	\end{center}	
\end{figure}

Figs.~\ref{fig3} -~\ref{fig6}  display the GDR spectra  in $ ^{128-164}\text{Sm} $ isotopes calculated with the four Skyrme forces, compared with the available experimental data \cite{carlos1974}. It needs to be pointed out that the experimental data for Sm isotopes from A=128 to A=142, and from A=152 to A=160, and $ ^{146}\text{Sm} $ are not yet available. The calculated GDR spectra in $ ^{144-154}\text{Sm}$ isotopes together with the available experimental data \cite{carlos1974} are shown in Fig.\ref{fig3}. It can be seen that all four Skyrme forces give generally acceptable agreement with the experiment with  a slight down-shift of the order of 0.5 MeV  for SLy5, SLy6 in the case of the spherical nucleus $^{144}\text{Sm}$ and the weakly deformed $^{148-150}\text{Sm}$ nuclei , and slight up-shift ($\sim$ 0.5 MeV) for SVbas force. The agreement is better for deformed $ ^{152-154}\text{Sm} $ nuclei , where all Skyrme forces produce the deformation splitting, in which rare-earth nuclei as Samarium (Sm) with neutron number N$\approx$90 show an example of shape transitions\cite{carlos1974, maruhn2005, benmenana2020}. For $^{144}\text{Sm}$ (N=82), its GDR strength has a single-humped shape. The vibrations along the three axes degenerate ,\textit{i.e}., they are the same resonance energy $E _{i}$  ($E _{x} =E _{y} =E _{z} $), which confirms that this nucleus is spherical due to the relation $E_{i} \propto R_{i}^{-1}$ where i$\in$\{x,y,z\} \cite{speth1981}. For $^{148}\text{Sm}$ and $^{150}\text{Sm}$ nuclei, the two resonance peaks move away slightly from each other but the total GDR presents one peak, so they are also weakly deformed nuclei with prolate shape. For $^{152}\text{Sm}$ and $^{154}\text{Sm}$ nuclei, the total GDR  splits into two distinct peaks which confirms that these nuclei are strongly deformed with prolate shape since the  oscillations along the major axis (K=0 mode) are characterized by  lower frequencies than the oscillations perpendicular to this axis (K=1 mode) \cite{speth1991}. 

The isotope $^{146}\text{Sm}$ for which we do not have experimental data, SLy6, Sly5 and SVbas give a weakly deformed nucleus ($\beta_{2}\simeq$0.06) where the resonance peaks along the major and the minor axis are very close together, whereas UNEDF1 gives an approximate spherical nucleus ($\beta_{2}\simeq$0.01). Calculations in Ref.~\cite{naz2018}, based on the self-consistent Relativistic-Hartree-Bogoliubov (RHB) formalism, predicted a shape coexistence for $^{146}\text{Sm}$. In order to verify the shape coexistence \cite{wood1992, heyde2011} in $^{146}\text{Sm}$ nucleus, we redid the static calculations several times from different initial deformations with SLy6 force. In all cases, we  obtained two minima (prolate and oblate) whose their properties are displayed in Table \ref{tab4}. We can see that the difference in energy between these two minima is around $\Delta$(B.E)$\simeq$ 0.07 MeV. This is a clear indication of a shape coexistence in $^{146}\text{Sm}$ nucleus. According to the value of deformation parameter $\gamma$, this competition of shape is between oblate ($\gamma= 60^\circ$) and prolate ($\gamma= 0^\circ$) shape, but the deformation is very weak ($\beta_{2}\simeq 0.05$) in both cases. Fig.\ref{fig4} shows the calculated GDR spectra corresponding to two minima (prolate, oblate). It confirms this suggestion: the upper panel (Fig.\ref{fig4}(a)) shows an oblate shape for $^{146}\text{Sm}$ due to oscillations along the shorter axis (K=0 mode) which are characterized by higher energies than the oscillations along the axis ($ \mid$K$\mid$=1 mode) perpendicular to it, while  the lower panel (Fig.\ref{fig4}(b)) shows a prolate shape for this nucleus. In both cases, the deformation splitting $\Delta$E between the two peaks is too small which confirms that this nucleus is very weakly deformed. 

 \begin{table}[!htbp]
 	\centering
 	\caption {The ground-state properties of two minima for $^{146}\text{Sm}$ nucleus.  \label{tab4}} 
 	{\begin{tabular}{@{}ccccccccc@{}} \hline
 			 Properties&& Prolate minimum &&& Oblate minimum&&&  \\
 			\hline
 			Binding energy (B.E) && -1999.73 MeV&&&  -1999.66 MeV&&& \\
 			Root mean square (r.m.s)  && 4.970 fm&&&  4.969 fm&&&  \\
 			Quadrupole deformation $\beta_{2}$  && 0.063 &&& 0.048&&&  \\
 			Deformation parameter $\gamma$ && $0^\circ$ &&& $60^\circ$&&&  \\
 			\hline
 	\end{tabular}}
 \end{table}
 
Fig.\ref{fig5} shows the GDR strength in neutron-deficient $^{128-142}\text{Sm}$ isotopes. We can see that the deformation decreases gradually from the well deformed nucleus $^{128}\text{Sm}$ ($\beta_{2}\simeq$0.4) to the approximate spherical one $^{142}\text{Sm}$ ($\beta_{2}\simeq$0.0) ,\textit{i.e}., when the neutron number N increases and closes to the magic number N=82. We note that all Skyrme forces in this work give almost the same GDR spectra except for $^{140}\text{Sm}$. According to the GDR strength along the three axes, the $^{128}\text{Sm}$ nucleus is weakly triaxial with SLy6, SLy5 and SVbas whereas it has a prolate shape with UNEDF1. For the $^{130-132}\text{Sm}$ isotopes, all the four Skyrme forces predict a prolate shape for them. For $^{134}\text{Sm}$, SVbas and UNEDF1 predict a prolate shape, while SLy5 and SLy6 give a weak triaxial shape. For $^{136-138}\text{Sm}$ isotopes, we can see that the oscillations along the three axes correspond to different resonance energies $E_{i}$ ($E_{x}\neq E_{y}\neq E_{z}$), which shows that these nuclei are deformed with triaxial shape. The four Skyrme forces give  different results for $ ^{140}\text{Sm} $ as displayed in Fig.\ref{fig5}. The SLy family (SLy5 and SLy6) predict a triaxial shape, SVbas predicts a prolate shape while UNEDF1 gives an approximate spherical shape. For  $ ^{142}\text{Sm} $, all Skyrme forces  predict a spherical shape where the GDR strengths along the three axes are identical ,\textit{i.e}., ($E_{x}= E_{y}= E_{z}$).

 Fig.\ref{fig6} shows the GDR strength in neutron-rich  $ ^{156-164}\text{Sm} $ isotopes. We can see that all Skyrme forces provide quite similar results. From $^{156}\text{Sm}$ (N=94) to $^{164}\text{Sm}$ (N=102), the deformation gradually gets broader, and their GDRs acquire a pronounced double-humped shape. Therefore, these nuclei are strongly deformed with prolate shape since the oscillations energies along the longer axis (z-axis) are lower than those of oscillations along the short axis (x and y axes) ,\textit{i.e}., $ E_{z}< E_{x}=E_{y} $.\\
 
\begin{figure}[!htbp]
	\centering
	\includegraphics[scale=0.7]{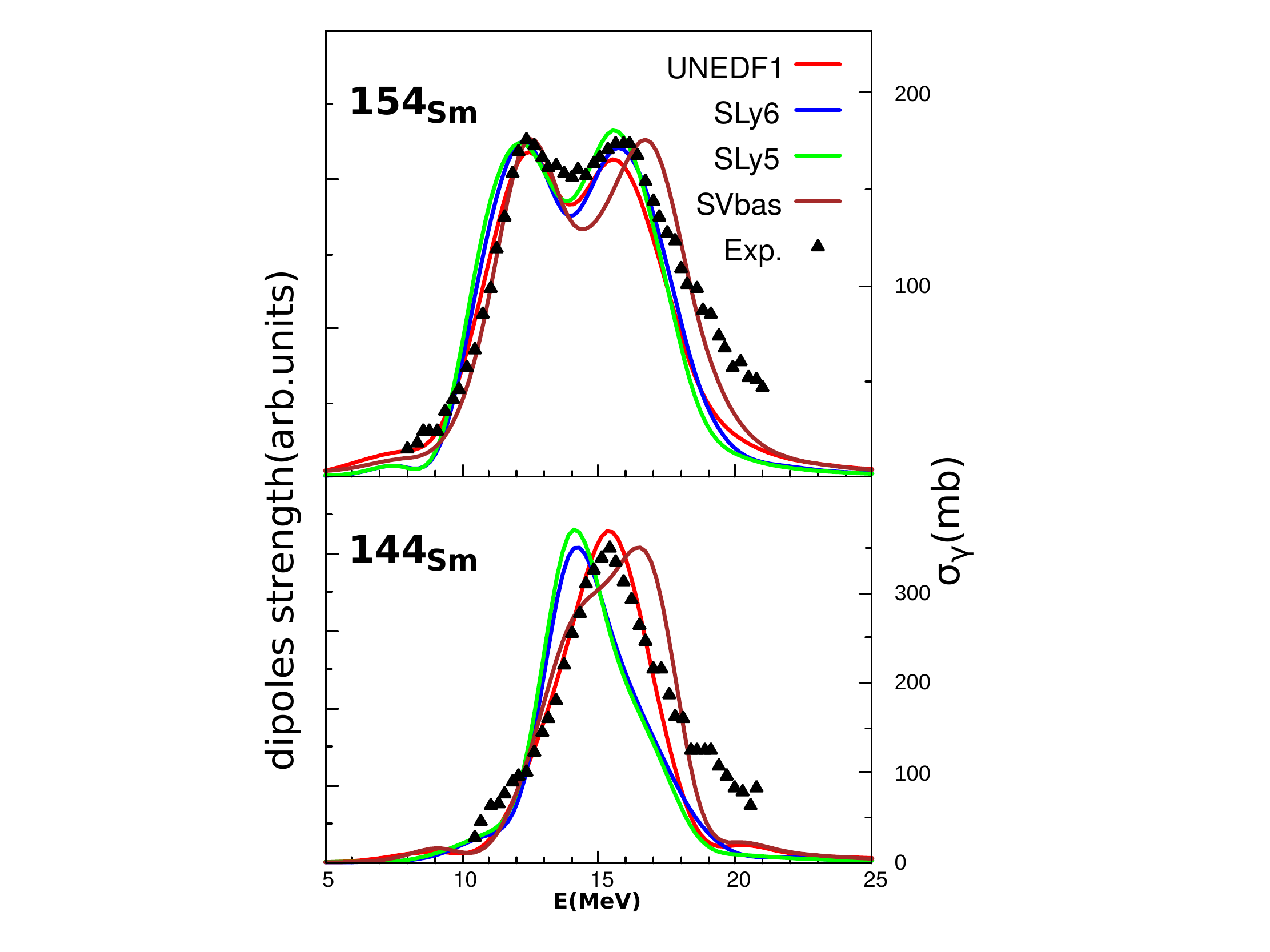}
	\caption{(Color online) The calculated GDR spectra $^{144}\text{Sm} $ and $^{154}\text{Sm} $ with Skyrme forces UNEDF1,  SLy6, SLy5 and SVbas for . the experimental data \cite{carlos1974}are  depicted by triangle.}
	\label{fig7}
\end{figure}

In order to compare the results between different Skyrme forces under consideration, we plot their GDR spectra into one figure, together with experimental data.
Fig.\ref{fig7} shows the GDR strength in $^{144}\text{Sm}$, and $ ^{154}\text{Sm} $ calculated by the four Skyrme forces as well as the experimental data from Ref.\cite{carlos1974}. It can be seen there is a dependence of the GDR spectra on various Skyrme forces. We note a small shift of the average peak position of  $\sim$ 1 MeV between these forces. The peak position of energy obtained with the Skyrme force SVbas is located highest among these four Skyrme forces. For the spherical nucleus $^{144}\text{Sm}$, the Skyrme force UNEDF1 reproduces well the shape and the peak among the four Skyrme forces. The agreement is less perfect with other forces. The SLy5 and SLy6 forces give very similar results, the strength exhibits a slight downshift while a slight upshift with SVbas functional. For the deformed nucleus $^{154}\text{Sm}$, there is an excellent agreement between the different functionals and the experiment, with a slight upshift for the K=1 mode for SVbas force. We can explain this dependence y the fact that it is linked to certain basic characteristics and nuclear properties of the Skyrme forces as shown in Table \ref{tab5}. The isovector effective mass $m_{1}^*/m$ is related to the sum rule enhancement  factor $\kappa$ by $m_{1}^*/m=1/(1+\kappa)$ \cite{berman1975}, \textit{i.e}., the larger isovector effective mass  corresponds to the lighter value of the enhancement factor. We can easily see that the increase of the  factor  $\kappa$ (\textit{i.e}., low isovector effective mass $ m_{1}^*/m $)  causes  the GDR strength to shift towards the higher energy region, as indicated in Ref.\cite{nesterenko2006} for the GDR in $^{154}\text{Sm}$, $^{238}\text{U}$ and $^{154}\text{No}$, and in Ref.\cite{oishi2016} for $ ^{174}\text{Yb}$. For example, the  large collective shift in SVbas can be related to a very high enhancement factor $\kappa$=0.4 compared to other Skyrme forces. In addition to the dependence with the enhancement factor $\kappa$, Fig.\ref{fig7} also shows a connection between GDR energy and symmetry energy $a_{sym}$. The peak energy of the GDR moves towards the higher energy region when $a_{sym}$ decreases, as pointed in Ref.\cite{stone2007} for the GDR in doubly magic $^{208}\text{Pb}$, and in our previous work for Nd isotopes \cite{benmenana2020}. 

\begin{table}[!htbp]
	\centering
	\caption {The sum rule enhancement  factor $\kappa$, isovector effective mass $m_{1}^*/m=1/(1+\kappa)$, and symmetry energy $ a_{sym} $ for the Skyrme forces under consideration.  \label{tab5}} 
	{\begin{tabular}{@{}ccccccccc@{}} \hline
			Forces && $m_{1}^*/m$ &&& $\kappa$&&& $ a_{sym}(MeV) $  \\
			\hline
			SLy6 \cite{CHABANAT1998} && 0.80&&&  0.25&&& 31.96\\
			SLy5 \cite{CHABANAT1998}  && 0.80&&&  0.25&&&32.03  \\
			UNEDF1~\cite{kortelainen2012}  && $\simeq$1.00 &&& 0.001&&&28.98  \\
			SVbas \cite{reinhard2009} && 0.715 &&& 0.4&&&30.00  \\
			\hline
	\end{tabular}}
\end{table}

\subsubsection{Relation between deformation splitting $\Delta E$ and quadrupole deformation $\beta_{2}$}
As we mentioned above, the GDR strength splits into two peaks for deformed nuclei. Each peak corresponds to a resonance energy $E _{i} $ of GDR. We denoted by $E_{1}$ and $E_{2}$ the energies  corresponding to K=0 and K=1 modes respectively. The total resonance energy of giant resonance is defined by the formula \cite{garg2018}
\begin{equation}{\label{eq17}}
E_{m} = \frac{\int_{0}^{+\infty} S(E)E dE}{\int_{0}^{+\infty} S(E) dE},
\end{equation}
where S(E) (\ref{eq12}) is the strength function of giant resonance.  In Table \ref{tab6} 
, the resonance energies E$_{1}$ and E$_{2}$ of $^{128-164}\text{Sm}$ nuclei are presented, including  the available experimental data from Ref.\cite{carlos1974}. From this table, we can see an overall agreement between our results and the experimental data, with a slightly advantage for the Sly6 functional. For instance, the result of the semi-spherical $^{144}\text{Sm} $ gives $ E_{GDR}^{SLy6} $=15.05 MeV which is very close to $ E_{GDR}^{Exp.} $=(15.30 $\pm$ 0.10) MeV. Also for deformed nuclei as $^{152}\text{Sm} $ and $^{154}\text{Sm} $, the results ($E_{1}, E_{2}$) with SLy6  are very close to  those of experiment.
\begin{table}[!htbp]
	\centering
	\caption {The resonance energy centroids $E_{1}$ and $E_{2}$ of  $^{128-164}\text{Sm}$ corresponding  to oscillation along the major axis (K=0)  and the minor axis (K=1) respectively. The experimental data are from ref. \cite{carlos1974}. \label{tab6}} 
	
	{\begin{tabular}{ccccccccccc}
			\cline{2-11}
			& \multicolumn{2}{c}{UNEDF1} & \multicolumn{2}{c}{SVBas}  &    \multicolumn{2}{c}{SLy5} &  \multicolumn{2}{c}{SLy6}& \multicolumn{2}{c}{Exp.\cite{carlos1971}} 	\\
			\hline Nuclei &    E$_{1}$ & E$_{2}$ &  E$_{1}$ & E$_{2}$ & E$_{1}$  & E$_{2}$ &  E$_{1}$& E$_{2}$ &  E$_{1}$ &  E$_{2}$	\\
			\hline	$^{\textbf{128}}\textbf{Sm}$ & 13.36 & 17.79 & 13.36 & 17.75 &  12.54 & 16.61 &12.76  & 16.90 & ---  &  ---	\\
			\hline	$^{\textbf{130}}\textbf{Sm}$ & 13.32 &17.68 &  13.46 & 17.64 &  12.58 & 16.49 & 12.82 & 16.76  & --- &  ---	\\
			\hline	$^{\textbf{132}}\textbf{Sm}$& 13.15 & 17.59 &  13.47 & 17.55 &  12.63 & 16.46  &12.89 & 16.69 & ---  &  ---	\\
			\hline	$^{\textbf{134}}\textbf{Sm}$ & 13.22 & 17.43 & 13.22 & 17.60 &  12.96 & 16.13 & 13.22  & 16.35 & ---  &  ---	\\
			\hline	$^{\textbf{136}}\textbf{Sm}$ & 14.00 &16.84 &  14.20 & 16.91 &  13.27 & 15.87 & 13.50 & 16.11  & --- &  ---	\\
			\hline	$^{\textbf{138}}\textbf{Sm}$& 14.34 & 16.51 &  14.47 & 16.66 &  13.48 & 15.70  &13.70 & 15.95 & ---  &  ---	\\
			\hline	$^{\textbf{140}}\textbf{Sm}$ & 15.42 & 15.73 & 14.93 & 16.25 &  13.73 & 15.50 & 13.95  & 15.72 & ---  &  ---	\\
			\hline	$^{\textbf{142}}\textbf{Sm}$ & 15.59 &15.59 &  15.78 & 15.78 &  14.80 & 14.83 & 15.03 & 15.04  & --- &  ---	\\
			\hline	$^{\textbf{144}}\textbf{Sm}$& 15.57 & 15.57 &  15.79 & 15.79 &  14.84 & 14.84  & 15.05 & 15.05 & 15.30$\pm$ 0.1  &  ---	\\
			\hline	$^{\textbf{146}}\textbf{Sm}$& 15.27 & 15.45 &  15.45 & 15.85  & 14.15 & 14.95  & 14.34 & 15.16  & --- & --- 	\\
			\hline	$^{\textbf{148}}\textbf{Sm}$ & 14.07 & 15.69 & 14.25 & 16.15 &  13.34 & 15.24  & 14.29 & 15.47 & 14.80$\pm$ 0.1  &  ---	\\
			\hline	$^{\textbf{150}}\textbf{Sm}$ & 13.40 & 15.86 & 13.62 & 16.31  & 12.91 & 15.38  & 13.06 & 15.59  & 14.60$\pm$ 0.1 &  ---	\\
			\hline	$^{\textbf{152}}\textbf{Sm}$ & 12.80 & 16.07 & 13.14 & 16.65  & 12.46  &15.62  & 12.60 & 15.82  & 12.45$\pm$ 0.1 &  15.85$\pm$ 0.1	\\
			\hline	$^{\textbf{154}}\textbf{Sm}$ & 12.53 & 16.06 & 12.93 & 16.98   &12.23 &15.70  &  12.37 & 15.91  & 12.35$\pm$ 0.1 & 16.10$\pm$ 0.1 \\
			\hline	$^{\textbf{156}}\textbf{Sm}$ & 12.36 & 15.97 & 12.80 & 16.55 &  12.12 & 15.64 & 12.26  & 15.82 & ---  &  ---	\\
			\hline	$^{\textbf{158}}\textbf{Sm}$ & 12.22 &15.84 &  12.69 &  16.47 & 12.01 & 15.60 & 12.15 & 15.77  & --- &  ---	\\
			\hline	$^{\textbf{160}}\textbf{Sm}$& 12.08 & 15.69 &  12.35 & 16.37 &  11.87 & 15.51  & 12.07 & 15.69 & ---  &  ---	\\
			\hline	$^{\textbf{162}}\textbf{Sm}$ & 11.96 & 15.53 & 12.52 & 16.26 &  11.87 & 15.40 & 11.99  & 15.57 & ---  &  ---	\\
			\hline	$^{\textbf{164}}\textbf{Sm}$ & 11.84 &15.37 &  12.44 &  16.14 & 11.82 & 15.33 & 11.95 & 15.50  & --- &  ---	\\
			\hline	
	\end{tabular}}
\end{table}

\begin{figure}[!htbp]
 	\centering
 	\includegraphics[scale=0.7]{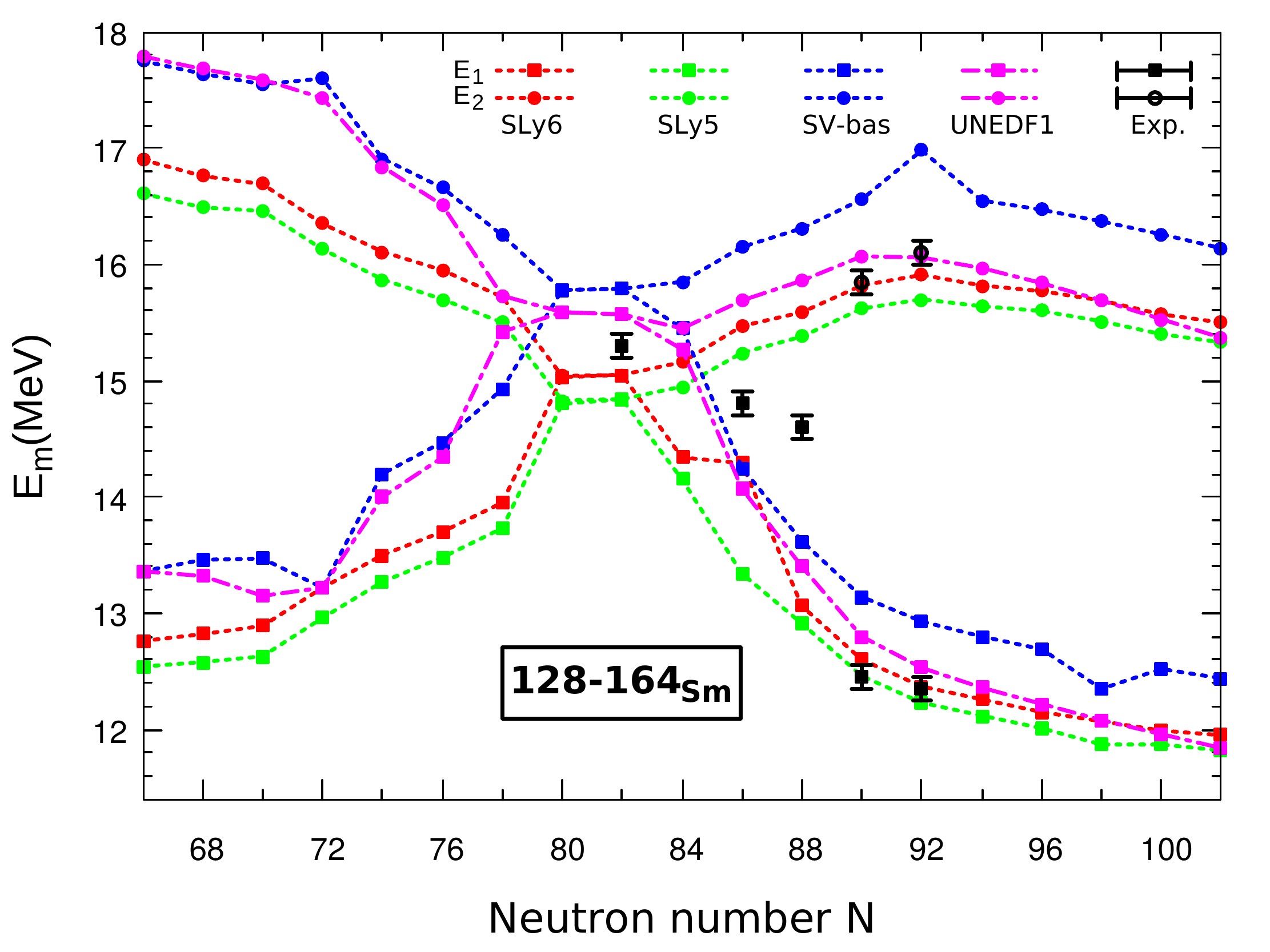}
 	\caption{ (Color online) The peak positions $E_{1}$ and $E_{2}$ of GDR in $^{128-164}\text{Sm} $ along major axis (square symbol) and minor axis (circle symbol) respectively. The experimental data are depicted by black square ($E_{1}$) and circle ($E_{2}$).}
 	\label{fig8}
 \end{figure}

 Fig.~\ref{fig8} displays the resonance energies ($E _{1}$, $E _{2}$) evolution as function of the neutron number N from $^{128}\text{Sm}$ (N=66) to $^{164}\text{Sm}$ (N=102). We can see for all the four Skyrme forces that  the resonance energy $E _{1}$ along the major axis (k=0 mode) increases with the neutron number N (\textit{i.e}., mass number A) until the region around N=82 (magic number) and then trends to decreases. The opposite happens for the resonance energy $E _{2}$, \textit{i.e}., it decreases with the increasing of N until N=82 , and then gradually increases. We can clearly see that the SLy6 reproduces the experimental data best among the four Skyrme forces. It was shown to provide a satisfying description of the GDR for spherical and deformed nuclei~\cite{nesterenko2008, reinhard2008}. The SVbas functional  gives somewhat high values of $E _{1}$ and $E _{2}$ among the other forces due to its large enhancement factor $\kappa$ ($\kappa$=0.4) as we discussed above.  

\begin{figure}[!htbp]
	\centering
	\includegraphics[scale=0.5]{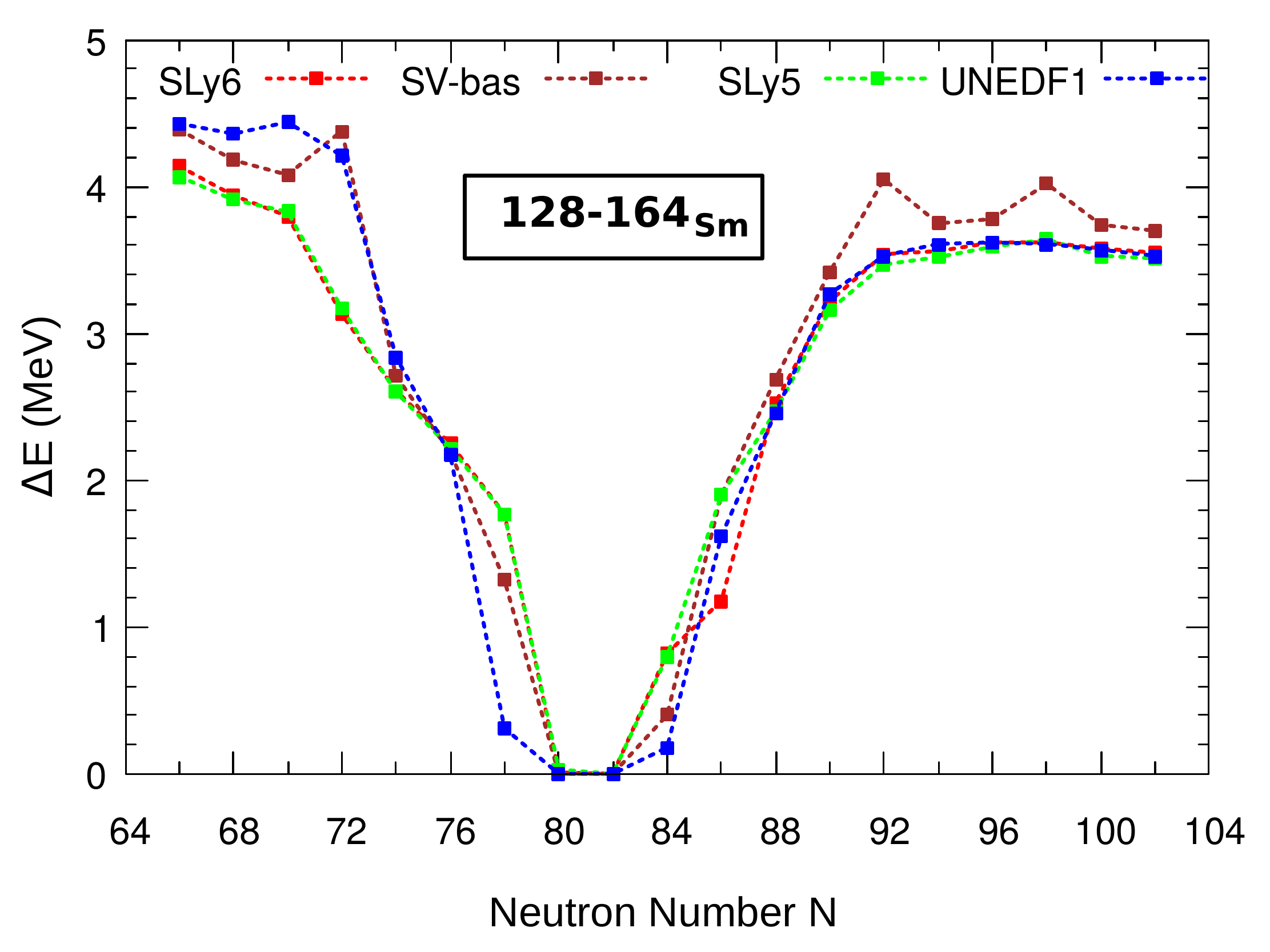}
	\caption{(Color online) The GDR-splitting $\Delta E$ as a function of the neutron number N for $ ^{128-164}\text{Sm} $ nuclei calculated with SLy6, SVbas, SLy5 and UNEDF1.}
	\label{fig9}
\end{figure}

In Fig.~\ref{fig9}, we plotted the evolution of the GDR-splitting value $\Delta E = E_{2} - E_{1}$ as a function of the neutron number N. It can be easily seen for all the four Skyrme forces, that the GDR splitting $\Delta$E decreases gradually with the increase of N and then increases.  It takes the minimum value $\Delta E$=0 at  N=82 (magic Number ) which corresponds to the spherical nucleus $ ^{144}\text{Sm} $ and achieves a maximum for
strongly deformed nuclei as $^{164}\text{Sm}$. Such a result confirms that the splitting of GDR is related to the deformation structure of nuclei.

\begin{figure}[!htbp]
	\begin{center}
		\begin{minipage}[t]{0.4\textwidth}
			\includegraphics[scale=0.3]{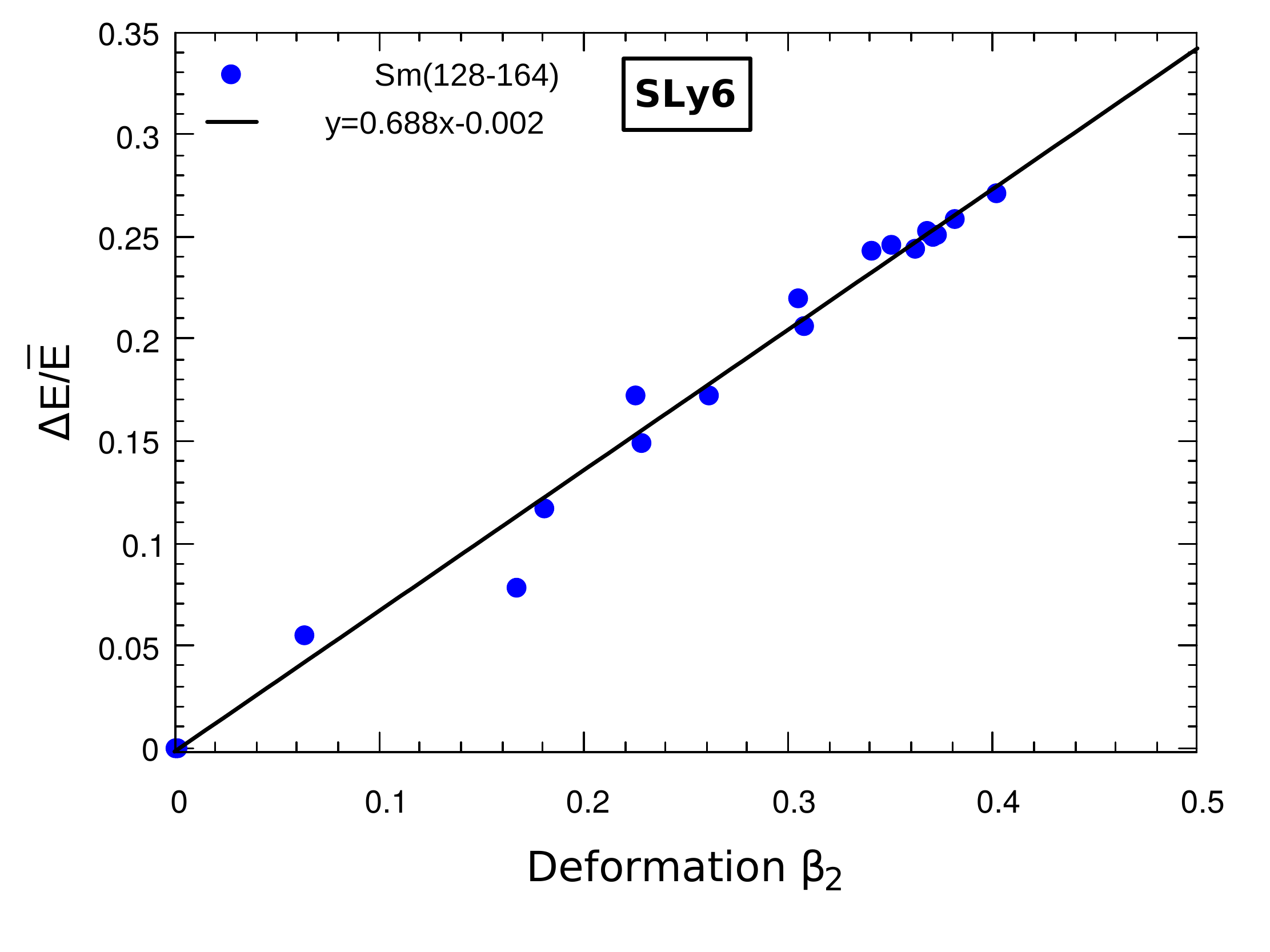}
		\end{minipage}
		\vspace{0.5cm}
		\hspace{2cm}
		\begin{minipage}[t]{0.4\textwidth}
			\includegraphics[scale=0.3]{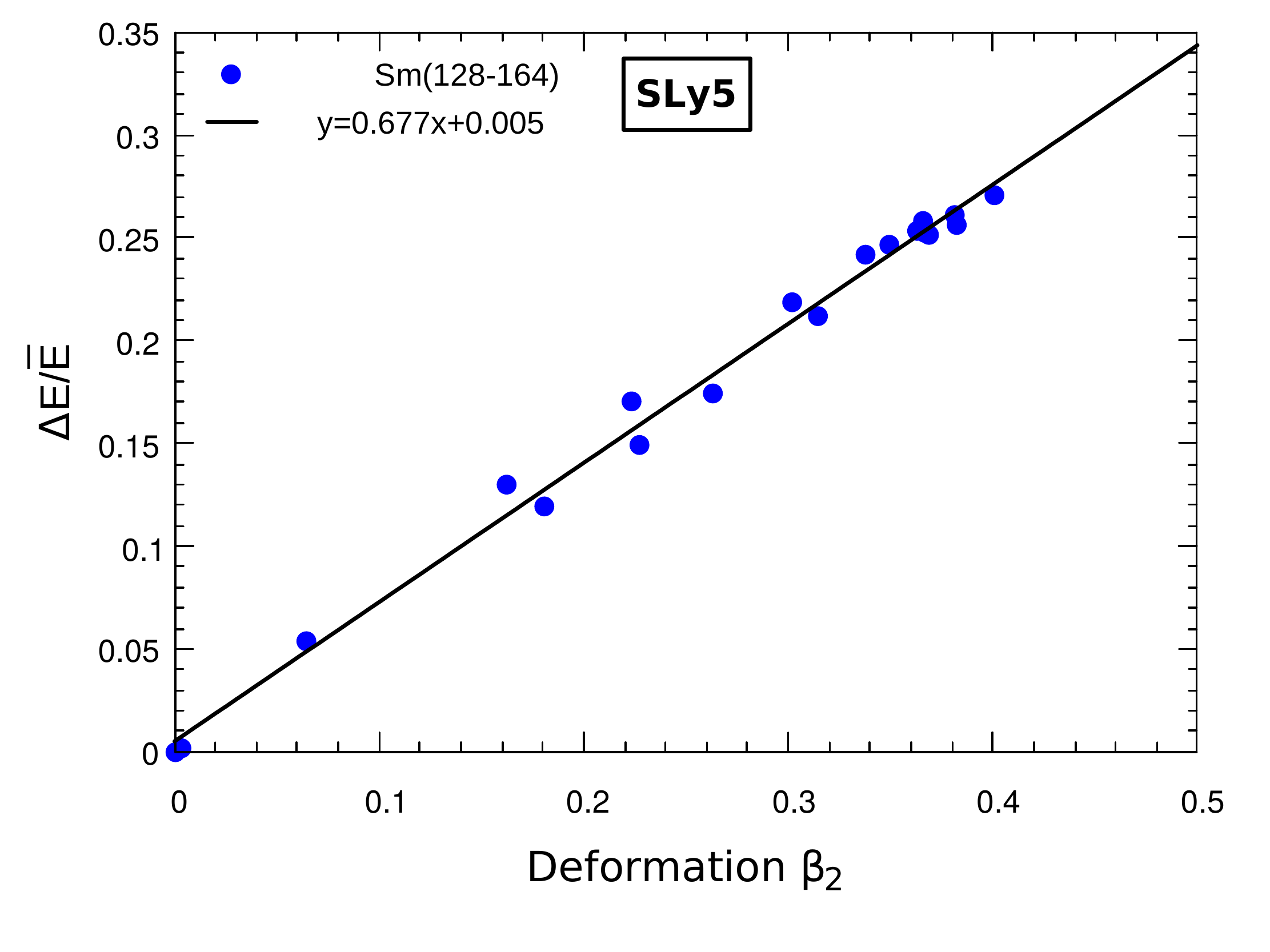}
		\end{minipage}
		\begin{minipage}[t]{0.4\textwidth}
			\includegraphics[scale=0.3]{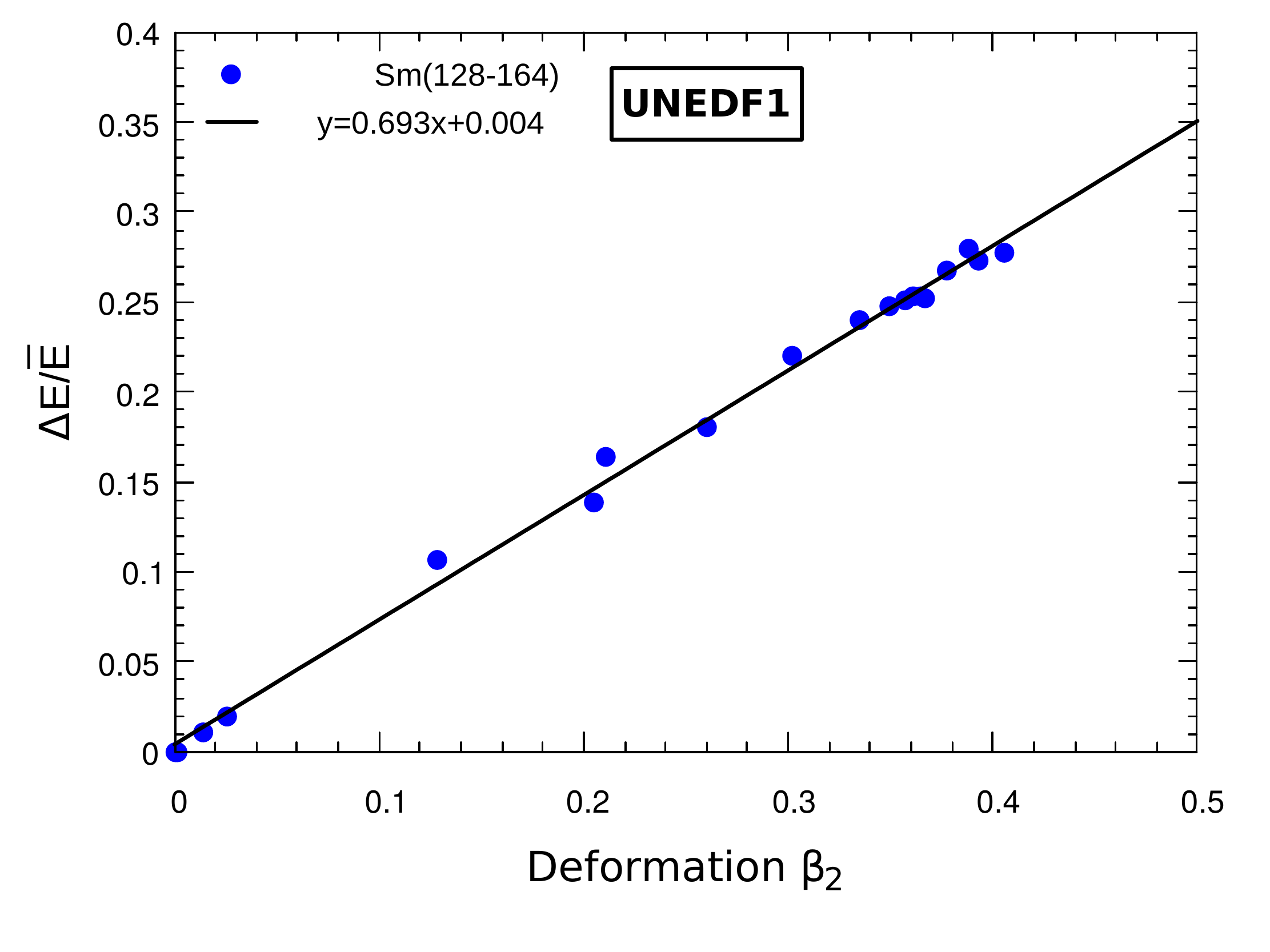}
		\end{minipage}
		\vspace{0.5cm}
		\hspace{2cm}
		\begin{minipage}[t]{0.4\textwidth}
			\includegraphics[scale=0.3]{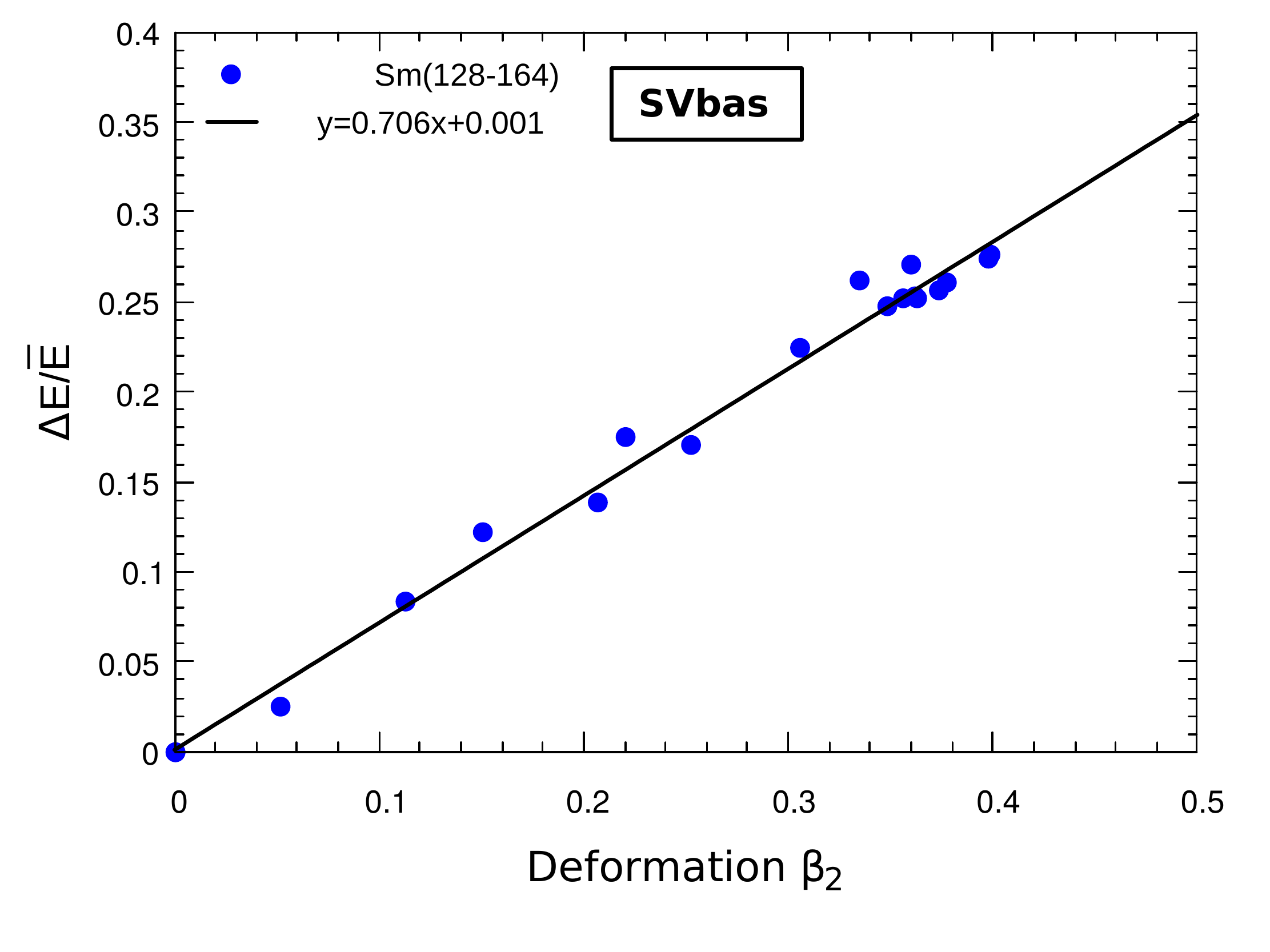}
		\end{minipage}
		\caption{(Color online) The correlation between the deformation parameter $\beta_{2}$ and the ratio $\Delta E/ \bar{E}$. circles denote the data in the  Sm isotopes and lines are the fitting results. } 
		\label{fig10}
	\end{center}	
\end{figure}

Since the GDR-splitting is caused by the deformation, it is possible to relate the nuclear deformation parameter $\beta_{2}$ with the ratio $\Delta E/\bar{E}$, where $\bar{E}$ is the mean resonance energy. Fig.~\ref{fig10} displays the correlation between the quadrupole deformation $\beta_{2}$ and  $\Delta E/\bar{E}$  for $ ^{128-164}\text{Sm}$ nuclei calculated with the Skyrme forces under consideration. We can see for all of the four Skyrme forces that there is an almost linear relationship between $\Delta E/\bar{E}$ and $\beta_{2}$, \textit{i.e}.,

\begin{equation}{\label{eq18}}
\Delta E/ \bar{E}  \simeq a.\beta_{2}, 
\end{equation}
where a is a parameter depending slightly on the Skyrme force. This fact confirms that the size of the GDR-splitting is proportional to the quadrupole deformation parameter $\beta_{2}$. The relation (\ref{eq18})  was already studied in Refs. \cite{okamoto1958,ring2009,benmenana2020}.

\section{Conclusion}\label{sec5}
\qquad The isovector giant dipole resonance (IVGDR) has been investigated in the isotopic chain of Samarium (Sm). The study covers  even-even Sm isotopes from $ ^{128}\text{Sm} $ to $ ^{164}\text{Sm}$. The investigations have been done within the framework of time dependent Hartree-Fock (TDHF) method based on the Skyrme functional. The calculations were performed with Four Skyrme forces: SLy6, SLy5, SVbas and UNEDF1. In static calculations, some properties of ground state like the deformation parameters ($\beta_{2}, \gamma$) have been calculated by using SKY3D code~\cite{sky3d}. In dynamic calculations, the dipole moment D$_{m}$(t) and the strength of GDR are calculated and compared with the available experimental data \cite{carlos1974}. The results obtained showed that TDHF method can reproduce the shape and the peak of the GDR spectra. All four Skyrme forces generally reproduce the average position of the GDR strength with a small shift depending on the used Skyrme force. The agreement is better with the SLy6 force among these Skyrme forces. The GDR strengths in  $ ^{128-142}\text{Sm} $, $ ^{146}\text{Sm} $ and $ ^{156-164}\text{Sm} $ nuclei are also predicted in this work.

Finally, some properties of GDR ($\bar{E}$, $E _{1} $, $E _{2} $, $\Delta E$ ) have been calculated with the four Skyrme forces. The results with SLy6 were very close to the experimental data compared to the other forces. A correlation between the ratio $\Delta E/ \bar{E}$ and the quadrupole deformation parameter $\beta_{2}$ was found. For all Skyrme forces, we have found the relation  $ \Delta E/ \bar{E}  = a.\beta_{2}+ b $ with the value of b is negligible.

In the light of the successful description of the GDR in deformed nuclei with the TDHF method, it was expected that this latter can also be applied for treating the shape coexistence as we predicted for 
$ ^{146}\text{Sm}$ with the SLy6 force.
\bibliographystyle{ieeetr}
\bibliography{sm.bib}

\begin{thebibliography}{10}

\bibitem{harakeh2001}
M.~N. Harakeh and A.~Woude, {\em Giant Resonances: fundamental high-frequency
  modes of nuclear excitation}, vol.~\textbf{24}.
\newblock Oxford University Press on Demand, (2001).

\bibitem{goldhaber1948}
M.~Goldhaber and E.~Teller {\em Phys. Rev.}, vol.~\textbf{74}, p.~1046, (1948).

\bibitem{speth1981}
J.~Speth and A.~van~der Woude {\em Reports on Progress in Physics},
  vol.~\textbf{44}, p.~719, (1981).

\bibitem{berman1975}
B.~L. Berman and S.~Fultz {\em Reviews of Modern Physics}, vol.~\textbf{47},
  p.~713, (1975).

\bibitem{bothe1937}
W.~Bothe and W.~Gentner {\em Z.Phys}, vol.~\textbf{106}, p.~236, (1937).

\bibitem{carlos1971}
P.~Carlos, H.~Beil, R.~Bergere, A.~Lepretre, and A.~Veyssiere {\em Nuclear
  Physics A}, vol.~\textbf{172}, p.~437, (1971).

\bibitem{carlos1974}
P.~Carlos, H.~Beil, R.~Bergere, A.~Lepretre, A.~De~Miniac, and A.~Veyssiere
  {\em Nuclear Physics A}, vol.~\textbf{225}, p.~171, (1974).

\bibitem{donaldson2018}
L.~Donaldson, C.~Bertulani, J.~Carter, and al. {\em Physics Letters B},
  vol.~\textbf{776}, p.~133, (2018).

\bibitem{goeke1982}
K.~Goeke and J.~Speth {\em Annual Review of Nuclear and Particle Science},
  vol.~\textbf{32}, p.~65, (1982).

\bibitem{maruhn2005}
J.~A. Maruhn, P.~G. Reinhard, P.~D. Stevenson, J.~R. Stone, and M.~R. Strayer
  {\em Phys. Rev. C}, vol.~\textbf{71}, p.~064328, (2005).

\bibitem{reinhard2008}
W.~Kleinig, V.~O. Nesterenko, J.~Kvasil, P.-G. Reinhard, and P.~Vesely {\em
  Phys. Rev. C}, vol.~\textbf{78}, p.~044313, (2008).

\bibitem{yoshida2011}
K.~Yoshida and T.~Nakatsukasa {\em Phys. Rev. C}, vol.~\textbf{83}, p.~021304,
  (2011).

\bibitem{benmenana2020}
A.~A.~B. Mennana, Y.~E. Bassem, and M.~Oulne {\em Physica Scripta},
  vol.~\textbf{95}, p.~065301, 2020.

\bibitem{speth1991}
S.~Josef, {\em Electric and magnetic giant resonances in nuclei},
  vol.~\textbf{7}.
\newblock World Scientific, (1991).

\bibitem{reinhard2007c}
V.~Nesterenko, W.~Kleinig, J.~Kvasil, P.~Vesely, and P.-G. Reinhard {\em
  International Journal of Modern Physics E}, vol.~\textbf{16}, p.~624, (2007).

\bibitem{fracasso2012}
S.~Fracasso, E.~B. Suckling, and P.~Stevenson {\em Physical Review C},
  vol.~\textbf{86}, p.~044303, (2012).

\bibitem{ring2009}
D.~P.~n. Arteaga, E.~Khan, and P.~Ring {\em Phys. Rev. C}, vol.~\textbf{79},
  p.~034311, (2009).

\bibitem{wang2017}
S.~S. Wang, Y.~G. Ma, X.~G. Cao, W.~B. He, H.~Y. Kong, and C.~W. Ma {\em Phys.
  Rev. C}, vol.~\textbf{95}, p.~054615, (2017).

\bibitem{Masur2006}
V.~M. Masur and L.~M. Mel'nikova {\em Physics of Particles and Nuclei},
  vol.~\textbf{37}, p.~923, (2006).

\bibitem{ramakrishnan1996}
E.~Ramakrishnan, T.~Baumann, and al. {\em Physical review letters},
  vol.~\textbf{76}, p.~2025, (1996).

\bibitem{gundlach1990}
J.~Gundlach, K.~Snover, J.~Behr, and al. {\em Physical review letters},
  vol.~\textbf{65}, p.~2523, (1990).

\bibitem{dirac1930}
P.~A.~M. Dirac {\em Mathematical Proceedings of the Cambridge Philosophical
  Society}, vol.~\textbf{26}, p.~376, (1930).

\bibitem{blocki1979}
J.~B{\l}ocki and H.~Flocard {\em Physics Letters B}, vol.~\textbf{85}, p.~163,
  (1979).

\bibitem{chomaz1987}
P.~Chomaz, N.~Van~Giai, and S.~Stringari {\em Physics Letters B},
  vol.~\textbf{189}, p.~375, (1987).

\bibitem{maruhn2006}
J.~A. Maruhn, P.-G. Reinhard, P.~D. Stevenson, and M.~R. Strayer {\em Phys.
  Rev. C}, vol.~\textbf{74}, p.~027601, (2006).

\bibitem{stevenson2004}
P.~Stevenson, M.~Strayer, J.~Rikovska~Stone, and W.~Newton {\em International
  Journal of Modern Physics E}, vol.~\textbf{13}, p.~181, (2004).

\bibitem{sky3d}
B.~Schuetrumpf, P.-G. Reinhard, P.~Stevenson, A.~Umar, and J.~Maruhn {\em
  Computer Physics Communications}, vol.~\textbf{229}, p.~211, (2018).

\bibitem{CHABANAT1998}
E.~Chabanat, P.~Bonche, P.~Haensel, J.~Meyer, and R.~Schaeffer {\em Nuclear
  Physics A}, vol.~\textbf{635}, p.~231, (1998).

\bibitem{reinhard2009}
P.~Kl\"upfel, P.-G. Reinhard, T.~J. B\"urvenich, and J.~A. Maruhn {\em Phys.
  Rev. C}, vol.~\textbf{79}, p.~034310, (2009).

\bibitem{kortelainen2012}
M.~Kortelainen, J.~McDonnell, W.~Nazarewicz, P.-G. Reinhard, J.~Sarich,
  N.~Schunck, M.~V. Stoitsov, and S.~M. Wild {\em Phys. Rev. C},
  vol.~\textbf{85}, p.~024304, (2012).

\bibitem{tao2013}
C.~Tao, Y.~Ma, G.~Zhang, X.~Cao, D.~Fang, H.~Wang, {\em et~al.} {\em Physical
  Review C}, vol.~\textbf{87}, no.~1, p.~014621, 2013.

\bibitem{paar2007}
N.~Paar, D.~Vretenar, E.~Khan, and G.~Colo {\em Reports on Progress in
  Physics}, vol.~\textbf{70}, no.~5, p.~691, 2007.

\bibitem{negele1982}
J.~W. Negele {\em Reviews of Modern Physics}, vol.~\textbf{54}, p.~913, (1982).

\bibitem{engel1975}
Y.~Engel, D.~Brink, K.~Goeke, S.~Krieger, and D.~Vautherin {\em Nuclear Physics
  A}, vol.~\textbf{249}, p.~215, (1975).

\bibitem{kerman1976}
A.~Kerman and S.~Koonin {\em Annals of Physics}, vol.~\textbf{100}, p.~332,
  (1976).

\bibitem{koonin1977}
S.~E. Koonin, K.~T.~R. Davies, V.~Maruhn-Rezwani, H.~Feldmeier, S.~J. Krieger,
  and J.~W. Negele {\em Phys. Rev. C}, vol.~\textbf{15}, p.~1359, (1977).

\bibitem{reinhard2007}
P.-G. Reinhard, L.~Guo, and J.~Maruhn {\em The European Physical Journal A},
  vol.~\textbf{32}, p.~19, (2007).

\bibitem{simenel2018}
C.~Simenel and A.~Umar {\em Progress in Particle and Nuclear Physics},
  vol.~\textbf{103}, p.~19, (2018).

\bibitem{simenel2012}
C.~Simenel {\em The European Physical Journal A}, vol.~\textbf{48}, p.~152,
  (2012).

\bibitem{flocard1978}
H.~Flocard, S.~E. Koonin, and M.~S. Weiss {\em Phys. Rev. C}, vol.~\textbf{17},
  p.~1682, (1978).

\bibitem{bonche1976}
P.~Bonche, S.~Koonin, and J.~W. Negele {\em Phys. Rev. C}, vol.~\textbf{13},
  p.~1226, (1976).

\bibitem{SKYRME1958}
T.~Skyrme {\em Nuclear Physics}, vol.~\textbf{9}, p.~615, (1958).

\bibitem{simenel2009}
C.~Simenel and P.~Chomaz {\em Phys. Rev. C}, vol.~\textbf{80}, p.~064309,
  (2009).

\bibitem{stevenson2008}
J.~M. Broomfield and P.~D. Stevenson {\em Journal of Physics G: Nuclear and
  Particle Physics}, vol.~\textbf{35}, p.~095102, (2008).

\bibitem{ring1980}
P.~Ring and P.~Schuck, {\em The nuclear many-body problem}.
\newblock Springer-Verlag, (1980).

\bibitem{reinhard2006}
P.-G. Reinhard, P.~D. Stevenson, D.~Almehed, J.~A. Maruhn, and M.~R. Strayer
  {\em Phys. Rev. E}, vol.~\textbf{73}, p.~036709, (2006).

\bibitem{meng2005}
J.~Meng, W.~Zhang, S.~Zhou, H.~Toki, and L.~Geng {\em The European Physical
  Journal A-Hadrons and Nuclei}, vol.~\textbf{25}, p.~23, (2005).

\bibitem{naz2018}
T.~Naz, G.~Bhat, S.~Jehangir, S.~Ahmad, and J.~Sheikh {\em Nuclear Physics A},
  vol.~\textbf{979}, p.~1, (2018).

\bibitem{takigawa2017}
K.~W. N.~Takigawa, {\em "Fundamentals of Nuclear Physics"}.
\newblock Springer Japan, (2017).

\bibitem{raman2001}
S.~RAMAN, C.~NESTOR, and P.~TIKKANEN {\em Atomic Data and Nuclear Data Tables},
  vol.~\textbf{78}, p.~1, (2001).

\bibitem{HFB}
J.-P. Delaroche, M.~Girod, J.~Libert, H.~Goutte, S.~Hilaire, S.~P{\'e}ru,
  N.~Pillet, and G.~Bertsch {\em Physical Review C}, vol.~\textbf{81},
  p.~014303, (2010).

\bibitem{moller2008}
P.~M{\"o}ller, R.~Bengtsson, B.~Carlsson, P.~Olivius, T.~Ichikawa, H.~Sagawa,
  and A.~Iwamoto {\em Atomic Data and Nuclear Data Tables}, vol.~\textbf{94},
  p.~758, (2008).

\bibitem{wood1992}
J.~Wood, K.~Heyde, W.~Nazarewicz, M.~Huyse, and P.~Van~Duppen {\em Physics
  reports}, vol.~\textbf{215}, p.~101, (1992).

\bibitem{heyde2011}
K.~Heyde and J.~L. Wood {\em Reviews of Modern Physics}, vol.~\textbf{83},
  p.~1467, (2011).

\bibitem{nesterenko2006}
V.~Nesterenko, W.~Kleinig, J.~Kvasil, P.~Vesely, P.-G. Reinhard, and D.~Dolci
  {\em Physical Review C}, vol.~\textbf{74}, p.~064306, (2006).

\bibitem{oishi2016}
T.~Oishi, M.~Kortelainen, and N.~Hinohara {\em Phys. Rev. C}, vol.~\textbf{93},
  p.~034329, (2016).

\bibitem{stone2007}
J.~R. Stone and P.-G. Reinhard {\em Progress in Particle and Nuclear Physics},
  vol.~\textbf{58}, p.~587, (2007).

\bibitem{garg2018}
U.~Garg and G.~Col{\`o} {\em Progress in Particle and Nuclear Physics},
  vol.~\textbf{101}, p.~55, (2018).

\bibitem{nesterenko2008}
V.~Nesterenko, W.~Kleinig, J.~Kvasil, P.~Vesely, and P.-G. Reinhard {\em
  International Journal of Modern Physics E}, vol.~\textbf{17}, p.~89, (2008).

\bibitem{okamoto1958}
K.~Okamoto {\em Phys. Rev.}, vol.~\textbf{110}, p.~143, (1958).

\end{thebibliography}

\end{document}